\documentclass[sigconf]{acmart}
\usepackage{bm}
\usepackage{slashbox,booktabs}
\usepackage{booktabs} 
\usepackage[caption=false,font=normalsize,labelfon
t=sf,textfont=sf]{subfig}

\setcopyright{none}

\renewcommand\footnotetextcopyrightpermission[1]{} 
\begin{document}
\title{A new PIR-based method for real-time tracking}

\author{Tianye Yang$^1$, Xuefeng Liu$^*$$^2$, Shaojie Tang$^3$, Jianwei Niu$^2$, Peng Guo$^1$}
\affiliation{$^1$ Huazhong University of Science and Technology, Wuhan, China}
\affiliation{$^2$ Beihang University, Beijing, China}
\affiliation{$^3$ The University of Texas at Dallas, Richardson, USA}
\thanks{Corresponding author Email: csxfliu@gmail.com}

\begin{abstract}	
Pyroelectric infrared (PIR) sensors are considered to be promising devices for device-free localization due to its advantages of low cost, less intrusive, and the immunity from multi-path fading. However, most of the existing PIR-based localization systems only utilize the binary information of PIR sensors and therefore require a large number of PIR sensors and a careful deployment. A few works directly map the raw data of PIR sensors to one's location using machine learning approaches. However, these approaches require to collect abundant training data and suffer from environmental change. In this paper, we propose a PIR-based device-free localization approach based on the raw data of PIR sensors. The key of this approach is to extract a new type of location information called as the azimuth change. The extraction of the azimuth change relies on the physical properties of PIR sensors. Therefore, no abundant training data are needed and the system is robust to environmental change.  Through experiments, we demonstrated that a device-free localization system incorporating the information of azimuth change outperforms the state-of-the-art approaches in terms of higher location accuracy. In addition, the information of the azimuth change can be easily integrated with other PIR-based localization systems to improve their localization accuracy.
\end{abstract}


\maketitle

\bibliographystyle{ACM-Reference-Format}
\bibliography{sample-bibliography}

\begin{thebibliography}{1}
	
	
\bibitem{cite:Vision-Based Human Tracking}
  Robert Bodor, Vision-Based Human Tracking and Activity Recognition, Proc. of the 11th Mediterranean Conf. on Control and Automation, 2003.
  
  \bibitem{cite:introduction of Pyroelectricity}
  Sidney B. Lang, Pyroelectricity: From Ancient Curiosity to Modern Imaging Tool, Physics today, 2005.
  
  \bibitem{cite:Technology for Human Detection and Counting with RF Sensor Networks}
  Shaufikah Shukri, Latifah Munirah Kamarudin, Device Free Localization Technology for Human Detection and Counting with RF Sensor Networks: A Review, Journal of Network and Computer Applications, 2017.
  
  \bibitem{cite:Taming the Inconsistency of Wi-Fi Fingerprints} Xi Chen, Chen Ma, Michel Allegue, and Xue Liu, Taming the Inconsistency of Wi-Fi Fingerprints for Device-Free Passive Indoor Localization, IEEE Conference on Computer Communications, 2017.
  
  \bibitem{cite:Accurate Location Tracking From CSI-Based Passive Device-Free Probabilistic Fingerprinting} Shuyu Shi, Stephan Sigg, Lin Chen, and Yusheng Ji, Accurate Location Tracking From CSI-Based Passive Device-Free Probabilistic Fingerprinting, IEEE TRANSACTIONS ON VEHICULAR TECHNOLOGY, VOL. 67, NO. 6, JUNE 2018.
  
  \bibitem{cite:Fresnel lens intro}
  Giuseppe A. Cirino, Robson Barcellos, Design, fabrication and characterization of Fresnel lens array with spatial filtering for passive infrared motion sensors, Proceedings of the International Society for Optical Engineering, 2006.
  
  \bibitem{cite:Wireless Infrared Communications} 
  JOSEPH M.KAHN, JOHN R. BARRY, Wireless Infrared Communications,Proceedings of the IEEE,1997.
  
  
  
  \bibitem{cite:Enhanced-Range Intrusion Detection}
  Sami A. Aldalahmeh, Amer M. Hamdan, Enhanced-Range Intrusion Detection Using Pyroelectric Infrared Sensors,Sensor Signal Processing for Defence, 2016.
  
  \bibitem{cite:Human Tracking With Wireless Distributed Pyroelectric Sensors}
  Qi Hao, David J. Brady,Human Tracking With Wireless Distributed Pyroelectric Sensors, IEEE SENSORS JOURNAL, VOL. 6, NO. 6, DECEMBER 2006.
  
  \bibitem{cite: Moving Targets Detection and Localization in Passive Infrared Sensor Networks}
  Zhiqiang Zhang, Xuebin Gao, Moving Targets Detection and Localization in Passive Infrared Sensor Networks, 10th International Conference on Information Fusion, 2007.
  
  \bibitem{cite:Distributed Multiple Human Tracking with Wireless Binary PIR Sensor Networks}
  Qi Hao, Fei Hu, Distributed Multiple Human Tracking with Wireless Binary Pyroelectric Infrared (PIR) Sensor Networks, IEEE SENSORS Conference, 2010.
  
  \bibitem{cite:Multiple-Target Tracking With Binary Proximity Sensors}
  JASPREET SINGH, RAJESH KUMAR, Multiple-Target Tracking With Binary Proximity Sensors, ACM Transactions on Sensor Networks, Vol. 8, No. 1, Article 5, 2011.
  
  \bibitem{cite:A novel multi-human location method}
  Bo Yang, Xiaoshan Li, A novel multi-human location method for distributed binary pyroelectric infrared sensor tracking system: Region partition using PNN and bearing-crossing location, Infrared Physics \& Technology, Volume 68, Pages 35-43, 2015.
  
  \bibitem{cite:Surveillance Tracking System Using}
  Byunghun Song , Haksoo Choi, Surveillance Tracking System Using Passive Infrared Motion Sensors in Wireless Sensor Network, International Conference on Information Networking, 2008.
  
  \bibitem{cite:The Smart-Condo}
  Iuliia Vlasenko, Ioanis Nikolaidis, Eleni Stroulia, The Smart-Condo: Optimizing Sensor Placement for Indoor Localization, IEEE TRANSACTIONS ON SYSTEMS, MAN, AND CYBERNETICS: SYSTEMS, VOL. 45, NO. 3, 2015.
  
  \bibitem{cite:Preprocessing Design in Pyroelectric}
  Jiang Lu, Ting Zhang, Fei Hu, Qi Hao, Preprocessing Design in Pyroelectric Infrared Sensor-Based Human-Tracking System: On Sensor Selection and Calibration, IEEE TRANSACTIONS ON SYSTEMS, MAN, AND CYBERNETICS: SYSTEMS, 2016.
  
  
  \bibitem{cite:MOLTS}
  Tao Liu, Yi Liu, Xiaozong Cui, Guangsheng Xu, Depei Qian, MOLTS: Mobile Object Localization and Tracking System Based on Wireless Sensor Networks, IEEE Seventh International Conference on Networking, Architecture, and Storage, 2012.
  
  \bibitem{cite:last binary work}
  Ibrahim Al-Naimi, Chi Biu Wong, Philip Moore, Advanced Approach for Indoor Identification and Tracking Using Smart Floor and Pyroelectric Infrared Sensors, International Conference on Information \& Communication Systems, 2014.
  
  \bibitem{cite:PIR Sensors Characterization and Novel Localization Technique}
  S. Narayana, R.V. Prasad, V.S. Rao, T.V. Prabhakar, PIR Sensors: Characterization and Novel Localization Technique, International Conference on Information Processing in Sensor Networks, 2015.
  
  \bibitem{cite:INDOOR USER ZONING AND TRACKING}
  Gianluca Monaci, Ashish Pandharipande, INDOOR USER ZONING AND TRACKING IN PASSIVE INFRARED SENSING SYSTEMS, 20th European Signal Processing Conference, 2012.
  
  
  
  \bibitem{cite:An Overview of particle filter}
  Olivier Cappé, Simon J. Godsill, and Eric Moulines, An Overview of Existing Methods and Recent Advances in Sequential Monte Carlo,  Proceedings of the IEEE ,Volume 95, Issue 5, 2007.
  
  \bibitem{cite:Modelling and Simulation of the Pyroelectric Detector}
  A. Odon, Modelling and Simulation of the Pyroelectric Detector Using MATLAB/Simulink, MEASUREMENT SCIENCE REVIEW, Volume 10, No. 6, 2010.
  
  
  \bibitem{cite:prominence of a peak}
  http://cn.mathworks.com/help/signal/ref/findpeaks.html\#buff2uu.
  
  
  \bibitem{cite:handbook of PIR sensors}
  Handbook of PIR sensors Tranesen-PCD-2F21:
  
  http://www.tranesen.com/portal.php?mod=list\&catid=1.
  
  
  
  \bibitem{cite:Handbook of Fresnel lens array}
  Handbook of Fresnel lens array YUYING-8719:
  
  http://www.keying-ly.com/upload/201410/24/201410241418320732.pdf.
  
  \bibitem{cite:inverse filter}
  M. Miyoshi, Y. Kaneda, Inverse filtering of room acoustics, IEEE Transactions on Acoustics, Speech, and Signal Processing, Volume: 36, Issue: 2, 1988.
  
  
  
  
  \bibitem{cite:pedestrian detection}
  Mykhaylo Andriluka, Stefan Roth, and Bernt Schiele, Pictorial Structures Revisited: People Detection and Articulated Pose Estimation, IEEE Conference on Computer Vision and Pattern Recognition, 2009.
%
	
\end{thebibliography}

\section{Introduction}
\label{sec:introduction}
Device-free localization has received much attention recently due to its advantage that the target does not need to carry any devices. Most device-free localization systems are either based on cameras \cite{cite:Vision-Based Human Tracking} or RF devices \cite{cite:Technology for Human Detection and Counting with RF Sensor Networks}\cite{cite:Taming the Inconsistency of Wi-Fi Fingerprints}\cite{cite:Accurate Location Tracking From CSI-Based Passive Device-Free Probabilistic Fingerprinting}. 
However, they both have important shortcomings. 
For example, visual-based localization systems may raise privacy concerns. On the other hand, systems based on RF signal generally suffer from multi-path effect, resulting unpredictable performances in practical conditions \cite{cite:Technology for Human Detection and Counting with RF Sensor Networks}. 

Recently, using pyroelectric infrared (PIR) sensors for device-free localization has started to attract much interest\cite{cite:introduction of Pyroelectricity}. PIR sensors are intrinsically immune to multi-path effect\cite{cite:Wireless Infrared Communications}, and therefore can be much more robust in a changing environment than those RF-based systems. As compared with camera based localization system, PIR sensors have much lower cost, with less privacy concerns, and can work well in a low-light environment. 

\begin{figure}[h]
\centering
\includegraphics[width=0.5\linewidth]{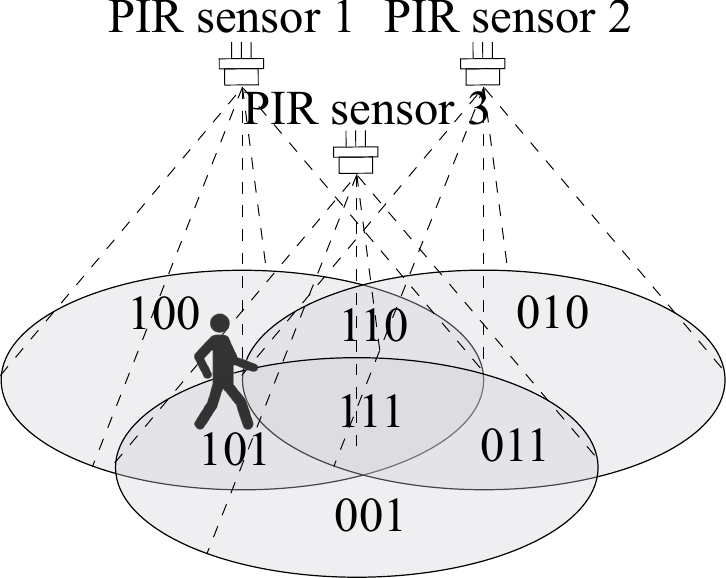}
\caption{Basic idea of PIR-based localization systems.}
\label{fig:Basic idea of BOB localization method}
\end{figure}

Most existing PIR-based localization systems \cite{cite:Human Tracking With Wireless Distributed Pyroelectric Sensors} -\cite{cite:last binary work} 
rely on the binary information extracted from the raw output of a PIR sensor. 
A PIR sensor generates binary information indicating the presence of a person in its detection zone. If we deploy a number of PIR sensors and make their detection zones partially overlap, the person at different locations will trigger different sets of PIR sensors. For example, the person in Fig. \ref{fig:Basic idea of BOB localization method} will trigger PIR sensor 1 and 3, and they will report '1'. While PIR sensor 2 will remain silent and report '0'. Therefore, the vector '101' can represent the current location. Similarly, vectors like '100','110','010' correspond to different locations shown in Fig. \ref{fig:Basic idea of BOB localization method}. By continuously collecting the vectors of these PIR sensors, we can obtain one's moving trajectory.

The localization accuracy of above approach is dependent on the granularity of the overlapped zones. The smaller the size, the higher the localization accuracy. As a consequence, it requires a lot of PIR sensors and a careful deployment to achieve high accuracy.

Note that a PIR sensor actually generates a voltage signal which reflects the amount of received infrared radiation. The voltage level is then compared to a pre-defined threshold and transforms to a binary value. Transforming an arbitrary voltage value to a binary one generally indicates a large amount of information loss.

Recently, some works start to explore the possibility of directly utilizing the raw voltage instead of binary information of PIR sensors. In \cite{cite:INDOOR USER ZONING AND TRACKING}, Gianluca Monaci et al. proposed an approach to estimate the information similar to the angle of arrival (AoA) from the raw data of PIR sensors. In \cite{cite:PIR Sensors Characterization and Novel Localization Technique},  the raw PIR data are utilized to estimate its distance to a person. Since much more information than the binary one is utilized, the number of PIR sensors required by a localization system is greatly reduced.

However, the above works still have some drawbacks. 
For example, their models are black-box and heavily rely on abundant of training data. In addition, models generally need to be re-trained in a new environment. 

In this paper, 
we develop a novel PIR-based localization system. 
It improves localization accuracy by utilizing a new type of information, called \textbf{azimuth change}, extracted from the raw data 
of PIR sensors. In addition, unlike the data-driven method proposed in \cite{cite:INDOOR USER ZONING AND TRACKING} and \cite{cite:PIR Sensors Characterization and Novel Localization Technique},  this information is mainly derived from the physical model of PIR sensors and therefore does not heavily rely on the labeled training data.

\begin{figure}[h]
\centering
\includegraphics[width=0.5\linewidth]{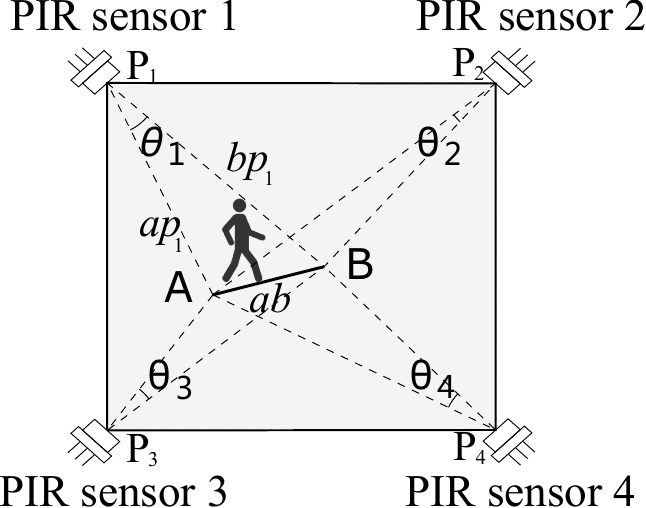}
\caption{The concept of azimuth change and how to determine one's location using the azimuth change.}
\label{fig:Basic idea of our localization method}
\end{figure}

The azimuth change is defined as the absolute difference of the azimuth of a person w.r.t a PIR sensor at two locations. For example, as shown in Fig.\ref{fig:Basic idea of our localization method}, when a person moves from A to B, the corresponding azimuth change w.r.t the 4 PIR sensors are $\theta _1$, $\theta _2$, $\theta _3$ and $\theta _4$, respectively. 

The azimuth change contains important information 
about one's location. 
Under some conditions, 
azimuth change alone 
would already be able to determine one's location. 
Taking Fig. \ref{fig:Basic idea of our localization method} as an example,
the location of the PIR sensor 1, i.e. $P_1$, along with two points $A$ and $B$ form a triangle. Using the cosine formula we can obtain the relationship among the azimuth change $\theta_1$ and the lengths of three sides of the triangle (i.e. $ap_1,bp_1$ and $ab$) as follows:
\begin{equation}
\cos {\theta _1} = \frac{{ap_1}^2 + {bp_1}^2 - {ab}^2}{2ap_1 \cdot bp_1}
\label{eq:eq1}
\end{equation}

Assume we have extracted the azimuth change $\theta_1$ from the raw data of PIR sensor 1. Eq. \ref{eq:eq1} contains three unknown lengths, i.e. $ap_1,bp_1,ab$. They can be further expressed by the coordinates of point $A$, $B$ and $P_1$. Considering the location of PIR sensor 1 is easily known, Eq. \ref{eq:eq1} only contains four unknown variables, i.e. the x-y coordinates of $A$ and $B$. Likewise, we can establish the similar equations w.r.t to other three PIR sensors. Correspondingly, we have a system of equations containing four equations and four variables. Theoretically speaking, the locations of both $A$ and $B$ are handily available by solving the system of equations.

\begin{figure}[h]
\centering
\includegraphics[width=0.8\linewidth]{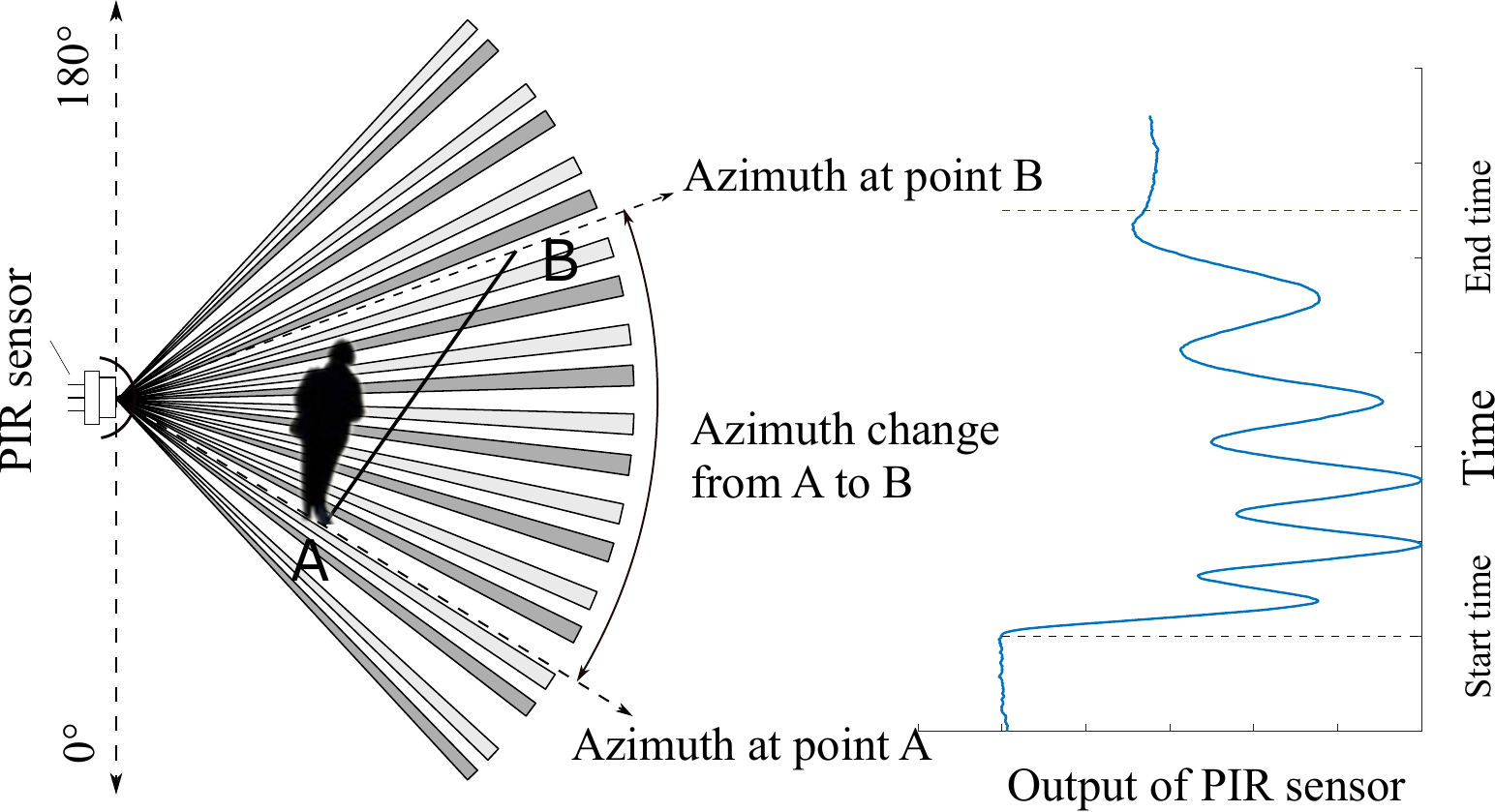}
\caption{Basic idea of our approach: the azimuth change of a moving person can be estimated by counting the number of rising and falling edges of the output of a PIR sensor.}%
\label{fig:sec1_PIROutput}
\end{figure}

The key problem then becomes how to extract the azimuth change from the raw data of PIR sensors. The approach is motivated by the following observations. In real conditions, a PIR sensor is always covered by a Fresnel lens array\cite{cite:Fresnel lens intro}, and the latter virtually divides the sensing zone of the PIR sensor into many evenly spaced fan-shaped zones, as shown in the left of Fig.\ref{fig:sec1_PIROutput}. Therefore, to know the azimuth change of the person when he moves from A to B, we only need know (1) the number of fan-shaped zones covered by the trace A-B, and (2) the azimuth change corresponding to two neighboring fan-shaped zones. With regard to identifying the first parameter, we will show that when a person moves from A to B, the output of a PIR sensor oscillates like a sine wave, as shown in the right of Fig.\ref{fig:sec1_PIROutput}. And the number of rising and falling edges of the sine wave equals to the number of the fan-shaped zones crossed. The second parameter can be obtained based on  the manual of PIR sensors. The details of estimating these two parameters will be described in Section \ref{sec:Limitations of estimating azimuth change directly by output}. 

Although the basic idea of estimating the azimuth change 
is not complicated, 
the implementation of this idea still entails 
several challenges.

For example, in some conditions, the output of PIR sensors changes slightly and the number of rising and falling edges in PIR sensors' output cannot be accurately estimated, especially in the presence of noise. This will result in an inaccurate estimation of azimuth change. By analyzing the physical property of PIR sensors, we found that using the PIR sensor's input inferred from the obtained output data can achieve much better performance. The details will be introduced in Section \ref{sec:Inaccuracy caused by extra rising and falling edges}. In addition, we found that when a person turn abruptly, the azimuth change cannot be estimated correctly. We proposed two methods to address this challenge and the details are described in Section \ref{sec:Influence of repeatedly crossed fan-shaped zones}.

The contributions of this paper are as follows:
\begin{itemize}
\item We find that from the output of a PIR sensor, we can extract the azimuth change of a moving person, which 
contains localization information.  To our best knowledge, this is the first time that this information is extracted and utilized for localization. 
\item We propose a set of methods to accurately estimate the azimuth change. In addition, these methods are  based on the PIR sensor's physical model and do not require users to collect abundant labeled training data.
\item We build a practical PIR-based localization system which achieves higher localization accuracy than the related state-of-art work, while the number of PIR sensors required by our system is only the half of theirs.
\end{itemize}



\section{The preliminary approach for estimating the azimuth change}
\label{sec:Limitations of estimating azimuth change directly by output}

\begin{figure}[h]
\centering
\includegraphics[width=0.35\linewidth]{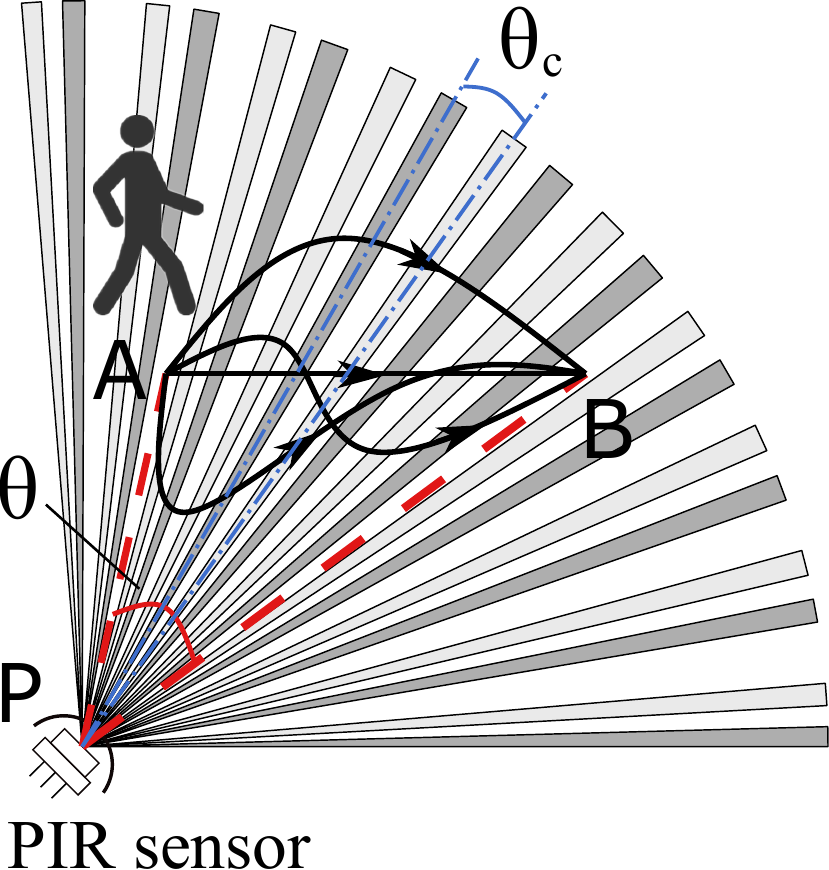}
\caption{The definition of the azimuth change.}
\label{fig:new Basic idea}
\end{figure}

We first give a formal definition of the azimuth change. As shown in Fig.\ref{fig:new Basic idea}, the \textbf{azimuth change} corresponding to the traces A-B w.r.t to the PIR sensor P, is defined as the included angle between the sides PA and PB, i.e. $\theta$. 

To obtain $\theta$, we rely on the fan-shaped zones shown in Fig.\ref{fig:new Basic idea}. But this leads to a natural question: where these fan-shaped zones come from? To answer the question, we need to understand the physical property of PIR sensors and the Fresnel lens array. 


\begin{figure}[h]%
\centering
\subfloat[][]{\centering
  \includegraphics[width=0.45\linewidth]{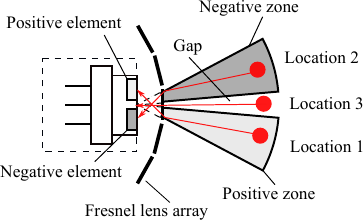}}%
\quad
\subfloat[][]{\centering
  \includegraphics[width=0.5\linewidth]{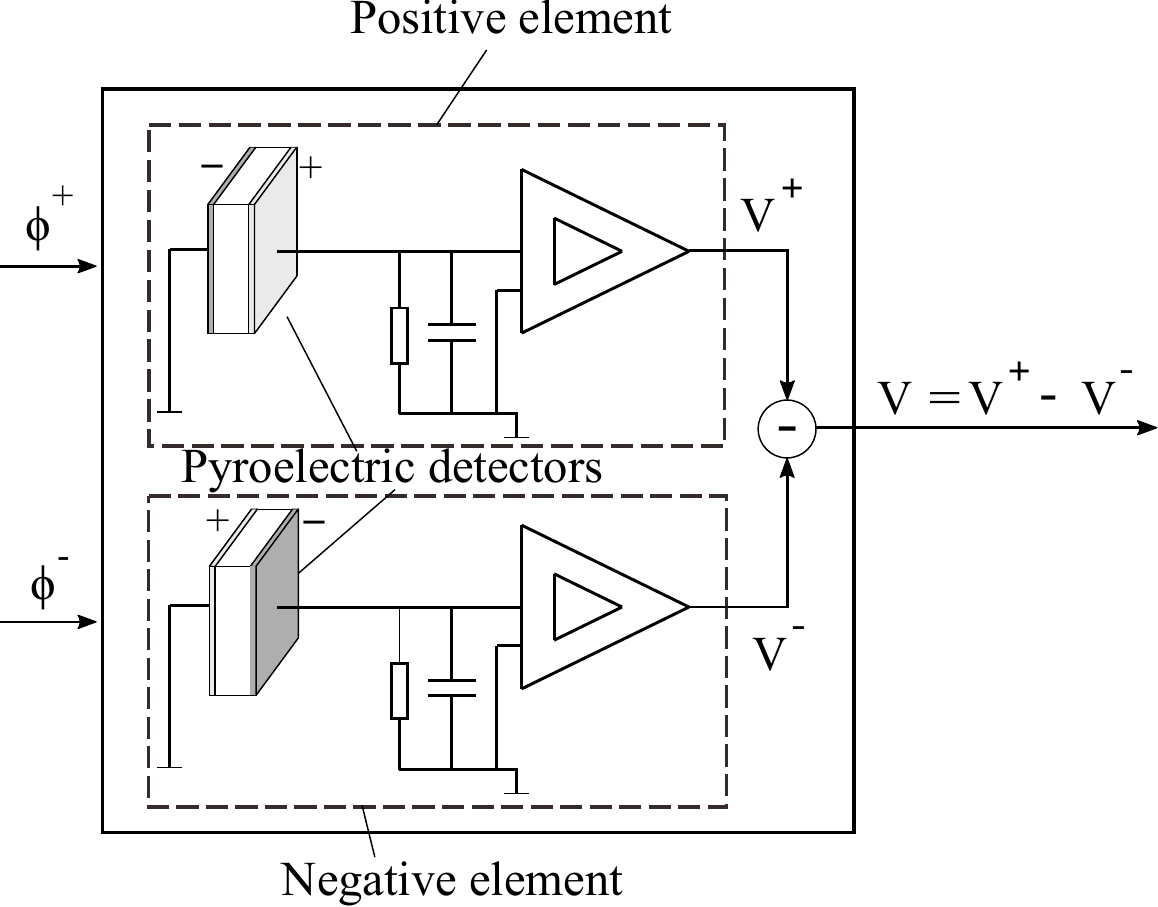}}
\caption[]{(a) The structure of PIR sensor and Fresnel lens, (b)the simplified internal structure of the sensing unit of a PIR sensor.}%
\label{fig:FunctionOfFresnelLens}
\end{figure}



Fig.\ref{fig:FunctionOfFresnelLens}(a) shows the structure of a typical PIR sensor covered by  Fresnel lens, and how it will be affected by an outside radiation source. Simply speaking, the infrared radiation of a moving person (denoted as the red dot) will first go through the Fresnel lens array, and then feed into the sensing unit which consists two parts, namely, a positive element and a negative one.  


Due to the structure of the Fresnel lens, when the heat source is at a certain location, say location 1 in the figure, only the positive element will receive the infrared radiation. Correspondingly, we say that location 1 is within a \emph{positive zone} of the PIR sensor. On the other hand, when the heat source is at location 2, only the negative element receives its infrared radiation, and we say that location 2 is in a \emph{negative zone} of the PIR sensor. Note that there exist locations as location 3, at which neither the positive nor negative element will receive the radiation. This indicates that there is a gap between the neighboring positive and negative zones.


Furthermore, the structure of Fresnel lens array \cite{cite:Handbook of Fresnel lens array} enables the following properties of the positive/negative zones: 
(1) both negative zones and positive ones are approximately fan-shaped zones centered at the PIR sensor, and with similar center angles. (2) There are many fan-shaped negative and positive zones, and they are alternatively placed, with one zone adjacent by two opposite ones. (3) The angles of the gaps between adjacent zones are similar. 

The above properties of fan-shaped zones can help us determine the azimuth change. More specifically,  as shown in Fig.\ref{fig:new Basic idea}, let $\theta_c$ be the included angle between the symmetric axes of two adjacent fan-shaped zones. Then the azimuth change $\theta$ can be approximately estimated by the equation as below:
\begin{equation}
\theta \approx N \cdot \theta_c
\label{eqn:EST by output}
\end{equation}
where $N$ is the total number of fan-shaped zones covered by the trace A-B. Our next task is to find out $N$ and $\theta_c$.



\textbf{To determine $N$}

We first show how to determine $N$. As shown in Fig. \ref{fig:sec1_PIROutput}, when a person moves in front of a PIR sensor, the PIR's output oscillates like a sine wave, this enables us to obtain $N$ by counting the number of rising and falling edges of the sine wave.

The rationale of above statement can be attributed to the physical properties of PIR sensors. Fig. \ref{fig:FunctionOfFresnelLens}(b) shows the simplified internal structure of a PIR sensor's sensing unit.



Let $\phi^+$ and $\phi^-$ be  the heat fluxes received by the positive and negative elements, respectively.  The two elements respectively generate an output voltage, denoted as $V^+$ and $V^-$.
The final output of the PIR sensor, denoted as $V$, is their difference:
\begin{equation}
V = V^+ - V^-
\end{equation}

When a person is moving from a positive zone to a negative one, $\phi^+$ keeps decreasing. Meanwhile,  $\phi^-$ keeps increasing. The above two effects combined to generate a decrease (i.e. a falling edge) in the PIR's output. Similarly, when the person is walking from a negative to a positive zone, we can expect to see an increase (i.e. rising edge) in the output $V$ of the PIR sensor.

As the positive zones and negative zones of PIR sensors are always alternatively placed to each other, we have an oscillating signal when a person is moving in front of a PIR sensor. Specifically, the total number of rising and falling edges is approximately equivalent to the number of zone he passes.

We utilize the following method to determine the number of rising and falling edges. First, we identify the peaks in the output signal of PIR sensors. Note that there are some peaks caused by environmental noise and should be eliminated. We utilize the feature called \textbf{prominence} \cite{cite:prominence of a peak}, which measures how much a peak stands out due to its intrinsic height and its location relative to other peaks. A peak with small prominence is regarded as noise and will be eliminated. Having determined the peaks, we can then calculate the number of rising and falling edges accordingly. 







\textbf{To determine $\theta_c$}


To estimate $\theta_c$, we first locate the radii of all fan-shaped zones, then calculate all the included angles between the symmetric axes of adjacent zones. At last, we  take the average as $\theta_c$.

Unfortunately, the locations of the radii of fan-shaped zones are generally not available in the manual of PIR sensors. Instead, a typical manual only provides the optical parameters about the Fresnel lens array (e.g. the focal length, width and direction of each Fresnel lens) and the PIR sensor (e.g. the width of positive/negative sensing units and their distances). 

\begin{figure}[h]
\centering
\includegraphics[width=0.9\linewidth]{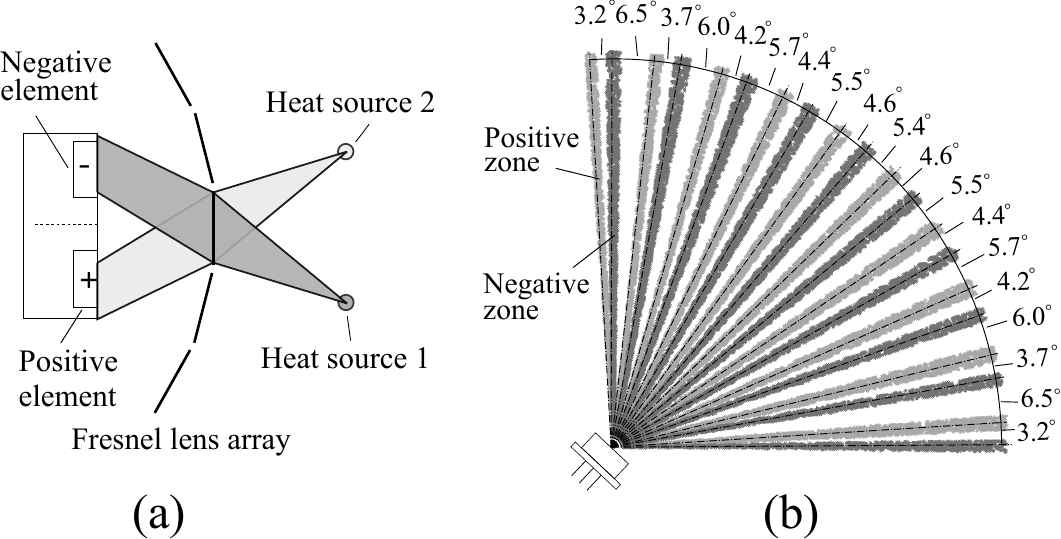}
\caption{Using the simulation method to determine $\theta_c$ of a PIR sensor. (a) How to color a simulated heat source, (b) The obtained layout of the PIR sensor.}
\label{fig:Where the emitted rays by a source will be focused}
\end{figure}

Correspondingly, we rely on these parameters and utilize a simulation approach to obtain $\theta_c$. The idea is similar to the 
Monte Carlo method. We generate a point heat source in the sensing range of the PIR sensor. Then based on the optical parameters of the Fresnel lens array, we calculate where the infrared radiation of the heat source will be focused on after passing through the Fresnel lens array. If the radiation is located on the negative element, then we color the heat source as `dark', indicating it is within a negative zone; if the radiation is focused on the positive element, the point heat source is colored as `gray', indicating it is within a positive zone. Otherwise, the heat source will be made transparent. As an example,  the point heat source 1 and 2 in Fig. \ref{fig:Where the emitted rays by a source will be focused} (a) are respectively colored as `dark' and `gray'. 

Then we generate a large number of point heat sources uniformly distributed within the sensing range of a PIR sensor. After all the points are colored, the boundaries of all the positive and negative zones emerge. The locations of the radii of all the positive and negative zones can be obtained by the boundaries of points with the same color. An example is shown in Fig. \ref{fig:Where the emitted rays by a source will be focused}(b).


Having identified the locations of all radii, we then find out the symmetric axis for each zone. The included angles of the symmetric axes between two neighboring zones are averaged to obtain $\theta_c$. Note that for three neighboring zones, like positive-negative-positive, the included angles of the first two and the last two are generally different. As shown in Fig. \ref{fig:Where the emitted rays by a source will be focused}(b), the included angle between the symmetric axes of the top positive zone and of the lower negative zone is 3.2$^\circ$, while the angle of the negative zone and lower positive one is 6.5$^\circ$. The average of all neighboring included angles shown in Fig. \ref{fig:Where the emitted rays by a source will be focused}(b), i.e. $\theta_c$, is $4.9^\circ$.  

We found through experiment that the simulation method described above can generate pretty accurate estimation of $\theta_c$. The detailed validation experiment and the results will be introduced in Section \ref{sec:testlayout}. 

\section{Advanced approach}

In the previous section, we introduced a preliminary design of our system. However, in some situations, direct utilizing this approach will lead to inaccurate estimation of $\theta$. In this section, we introduce two major enhancements. The first enhancement is to improve the accuracy of $N$, and second is to eliminate the negative effect of abrupt change of directions.

\subsection{Enhancement 1: Utilizing input instead of output of PIR sensor}
\label{sec:Inaccuracy caused by extra rising and falling edges}

In the previous section, we demonstrate that $N$ (the number of zones corresponding to the azimuth change) can be estimated by counting the number of rising/falling edges in the output data of PIR sensor. However, this approach may not give accurate estimation of $N$, especially in the following two conditions.

\begin{figure}[h]
\centering
\includegraphics[width=0.99\linewidth]{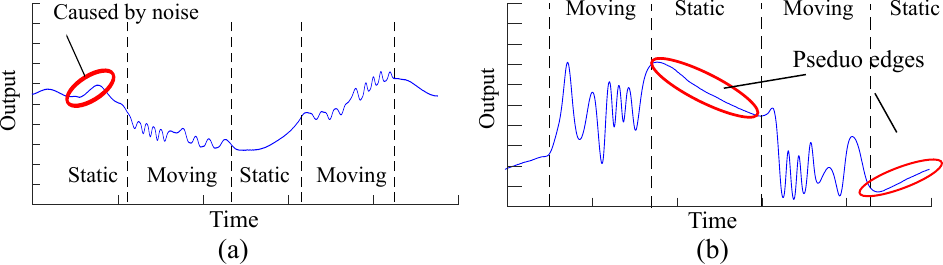}
\caption{Two typical scenarios when $N$ cannot be estimated accurately. (a) When a person moves fast, (b) when a moving person suddenly stops.}%
\label{fig:LimitationOfOutput}
\end{figure}

First, when a person moves relatively fast, being far away from a PIR sensor, or move parallel to the zones of a PIR sensor, the amplitude of the PIR's output is relatively small. Considering the effect of noise, it is not always easy to distinguish whether a rising or falling edge is caused by the person's movement or by noise, which generates error in $N$. An example is shown in Fig. \ref{fig:LimitationOfOutput}(a), where the output data of a PIR sensor are illustrated. The data were collected when a person was moving at about 2m/s and about 5 meters away from the PIR sensor. We can see that the PIR sensor's output oscillates slightly, and its amplitude is comparable to that caused by noise when the person is static. Correspondingly, it is difficult to obtain a proper threshold to distinguish a real rising/falling edge caused by movement and a pseudo one caused by noise, thus introducing error in $N$. 

Second, we found through experiment that each time a moving person stops, a PIR sensor will generate some pseudo rising or falling edges. An example is shown in Fig.\ref{fig:LimitationOfOutput}(b). These pseudo edges will also cause an inaccurate estimate of $N$.

\begin{figure}[h]
\centering
\includegraphics[width=0.5\linewidth]{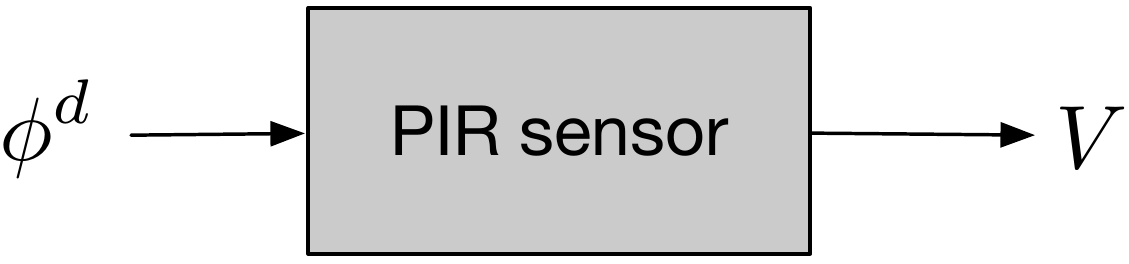}
\caption{A PIR sensor can be regarded as a dynamic system.}
\label{fig:PIR_model}
\end{figure}

To solve the above problems, we need to explore deeper into the principle of PIR sensors. We have shown in Fig.\ref{fig:FunctionOfFresnelLens} the internal structure of a PIR sensor. Considering the symmetry of the positive and negative elements, the effect of two inputs (i.e. heat fluxes) of PIR sensor, $\phi^+$ and $\phi^-$, can be replaced by their difference, i.e. $\phi^d = \phi^+ -\phi^-$. Thus, the PIR sensor shown in Fig. \ref{fig:FunctionOfFresnelLens} can be approximately represented as a dynamic system as shown in Fig. \ref{fig:PIR_model}.

The system's input $\phi^d$ is the \underline{d}ifference of the \underline{h}eat \underline{f}luxes, and is abbreviated as \textbf{DHF}. The output of the system is the original PIR sensor's output $V$.  The system can be characterized as a second-order dynamic system as follows\cite{cite:Modelling and Simulation of the Pyroelectric Detector}: 
\begin{equation}
{G}(s) = \frac{{V(s)}}{{{\phi ^d}(s)}} = \frac{As}{{B{s^2} + Cs + 1}}
\label{eqn:transform function of dual-element PIR sensor}
\end{equation}
where the $G(s)$ is the transfer function of the PIR sensor; $V(s)$ and $\phi^d(s)$ are the Laplace transform of $V(t)$ and $\phi^d(t)$, respectively. The parameters $A,B,C$ are determined by the internal electric circuits of the PIR sensor. 

Generally speaking, the system described in Eq.\ref{eqn:transform function of dual-element PIR sensor}  can be regarded as a low-pass filter. In other words, \emph{the PIR sensor's output is essentially a filtered version of the DHF}. 

\begin{figure}[h]
\centering
\includegraphics[width=0.8\linewidth]{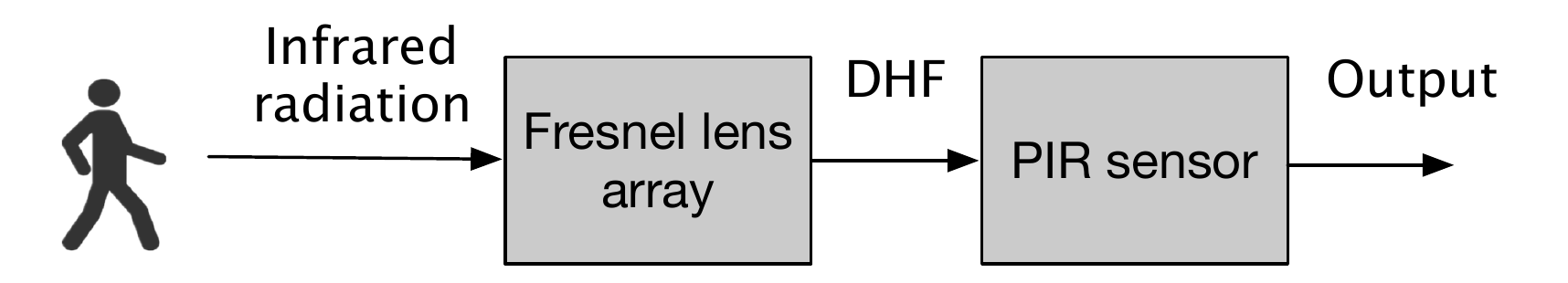}
\caption{How a person's infrared radiation affects the output of a PIR sensor.}
\label{fig:three system}
\end{figure}

Furthermore, we can see from a broader perspective how a person's infrared radiation will affect the output of a PIR sensor. As shown in Fig. \ref{fig:three system}, when a person is moving in front of a PIR sensor, the person's infrared radiation will first pass through the Fresnel lens array to generate the DHF, which passes through the system of PIR sensor, and finally generates the output. 

From system point of view, the first unit in Fig. \ref{fig:three system}, i.e. Fresnel lens array, can be described as a linear system without delay. This is because $\phi^+$ and $\phi^-$ at a certain time is directly determined by the area of the person's body in the corresponding zones at the same time\cite{cite:Enhanced-Range Intrusion Detection}. Furthermore, when the person is in the center of a positive zone, the DHF will have a maximum value (i.e. a peak). Similarly, a person at the center of a negative zone will generate a minimum value (i.e. a trough). 
In contrast, the second unit in Fig. \ref{fig:three system}, i.e. PIR sensor, is a low-pass filter which inevitably introduces attenuation, smoothing and delay effect to the output.

The above discussion naturally leads to a conclusion: \textbf{using DHF instead of the output of PIR sensors can potentially give a better estimation of $N$}, as the former is more directly affected by the person's infrared radiation than the latter.   

\begin{figure}[h]
\centering
\includegraphics[width=0.99\linewidth]{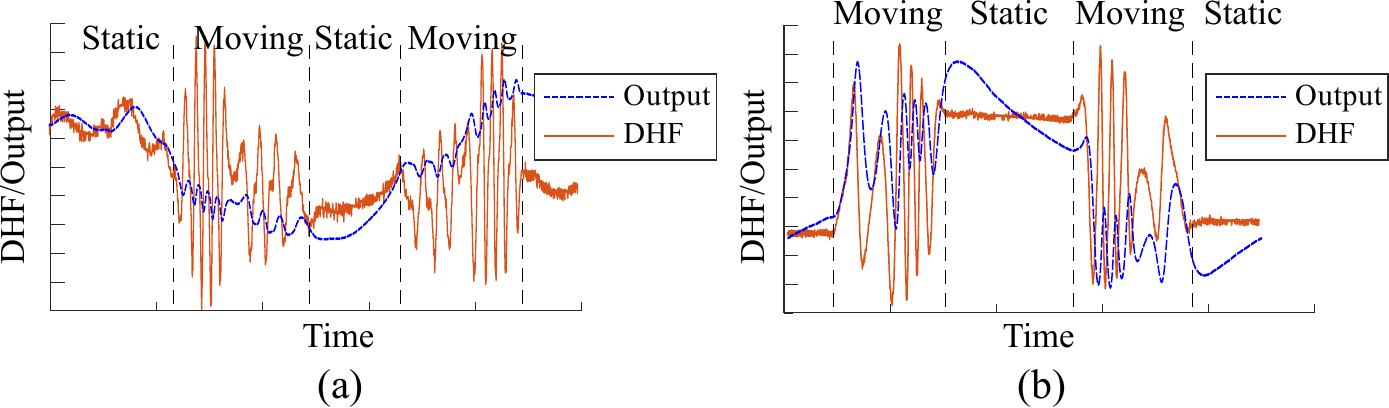}
\caption{Using DHF instead of the output of PIR sensor can improve the accuracy of $N$.}%
\label{fig:DHF}
\end{figure}

For comparison, Fig. \ref{fig:DHF}(a) show the DHF data collected in the experiment in which the output data have been shown in Fig.\ref{fig:LimitationOfOutput}(a). We can see that in the DHF data, the rising and falling edges caused by the person's movement are much more significant compared to those in the output of the PIR sensor. This indicates that we can easily find a good threshold of the prominence level that can well distinguish edges caused by movement and noise, i.e. a better estimation of $N$.  In addition, Fig. \ref{fig:DHF}(b) shows that there is no pseudo edges in the DHF.

In the remaining of this section, we briefly introduce how to estimate the DHF. Given the output of a PIR sensor, to estimate its DHF in essence, is an inverse problem and can be solved by a technique called as inverse filter \cite{cite:inverse filter}. 

Designing an inverse filter further requires the system parameters, i.e. the $A,B,C$ in Eq. \ref{eqn:transform function of dual-element PIR sensor}. To identify these three parameters, we design a simple experiment that generating a step input to the PIR sensor, and then measure system's output, which is generally called as step response in control engineering. According to the Eq. \ref{eqn:transform function of dual-element PIR sensor}, the theoretical step response can be represented as a function of $A,B,C$. Then these three parameters can be estimated using some optimization method similar to curve fitting.


The experiment to generate the step input is as follows. First, we cover the PIR sensor with a box that can isolate outside infrared radiation. Then we heat a soldering iron  to a constant temperature of $250^{\circ}$ and move it near to the PIR sensor with the distance about 0.5 meter. At last, we remove the box promptly. The last procedure will generate an abrupt constant input, i.e. a step, to the PIR sensor. We repeat the above experiment 10 times, and all the collected data are combined to estimate the corresponding parameters. Having identified the system parameters, we then utilize the inverse filtering technique to estimate the DHF from the output of PIR sensor.

\subsection{Enhancement 2: Handling abrupt turning}
\label{sec:Influence of repeatedly crossed fan-shaped zones}

\begin{figure}[h]%
\centering
\subfloat[][]{\centering
	\includegraphics[width=0.3\linewidth]{Fig48MovingAlongCurves.pdf}}\qquad 
\subfloat[][]{\centering
	\includegraphics[width=0.3\linewidth]{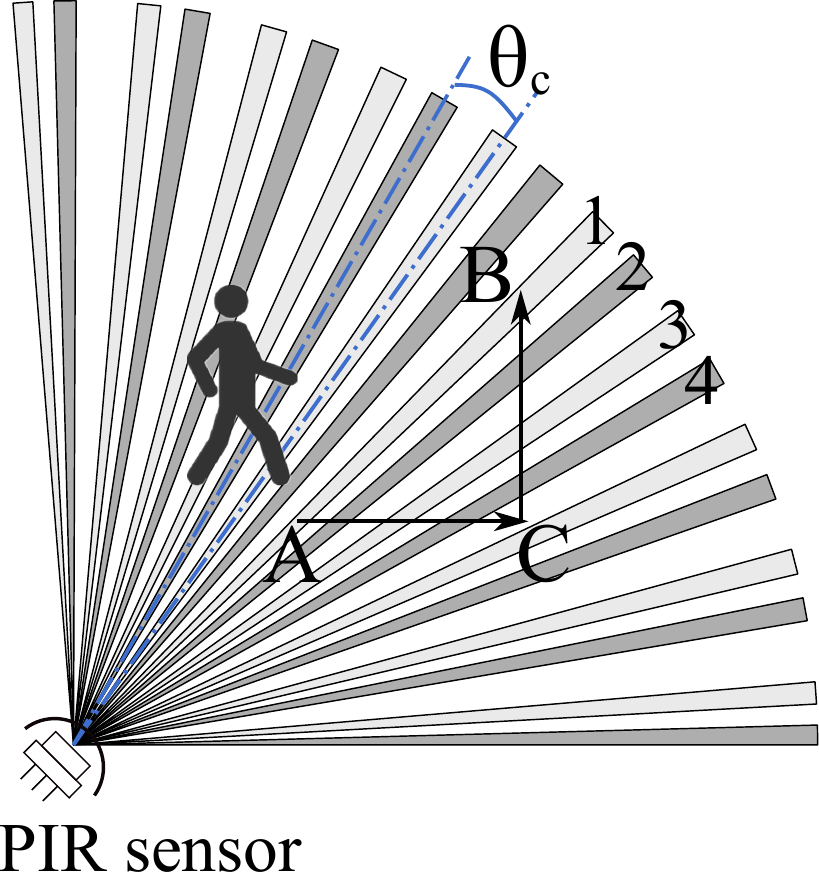}}
\caption[]{(a) Conditions when $N$ can be estimated accurately. (b) The previous method will fail if the trace contains abrupt turning points. }%
\label{fig:LimitationOfOutput2}
\end{figure}

We have shown in Section \ref{sec:Limitations of estimating azimuth change directly by output} that even when a person is walking on a curved trace (as those shown in Fig. \ref{fig:LimitationOfOutput2}(a)), the azimuth change can be estimated accurately by the product of the number of zones he crossed (i.e. $N$) and $\theta_c$. In this example, the azimuth change corresponding to curves A-B is $\theta= 8\cdot \theta_c$.

However, things will be different in Fig.\ref{fig:LimitationOfOutput2}(b). On the curve of A-B, the person has crossed a total of 8 zones (four during A-C and four during C-B). Correspondingly, the azimuth change $\theta=8\cdot \theta_c$. However, as can be easily seen from Fig.\ref{fig:LimitationOfOutput2}(b), the true azimuth change of trace A-B is  $0^\circ$. 

The key difference between the traces shown in Fig. \ref{fig:LimitationOfOutput2}(a) and Fig. \ref{fig:LimitationOfOutput2}(b) lies in the change of direction during A-B. In the former, the person's walking direction remains to be clockwise from A-B. In contrast, the direction of the curve in the latter starts from clockwise (A-C) but then changes to anticlockwise (C-B) at point C. For convenience, we define points where the direction has changed from clockwise to anticlockwise (or vice-versa) as the \textbf{turning points}. The presence of turning points makes the previous azimuth change estimation method fail. 


A straightforward approach to handle the problem is to decrease the estimation period.  Still use example shown in Fig.\ref{fig:LimitationOfOutput2}(b). Instead of estimating the azimuth change corresponding to the whole trace A-B, we can adopt a shorter estimation period, for example, a period that can divide A-B into two sections: A-C and C-B. In this case, the azimuth change during A-C and C-B are estimated accurately and no error will be introduced.

In real conditions, we generally cannot totally eliminate the error by reducing the estimation period as the example above. However, the probability of the presence of turning points and their effect will be lower in a shorter estimation period than in a longer one.

It should be noted that, the duration of the estimation period cannot be set to be too short.  If a person is walking as in Fig.\ref{fig:LimitationOfOutput2}(a), a longer estimation period normally can help to give better estimate of $\theta$ as it contains more information. In addition, an estimation period should be long enough to contain at least one or more crossed zones to give a good estimate of $N$. Through experiments, we set the estimation period to 0.5s for our system. 

Considering the downside of reducing the estimation period, another approach is to first detect these turning points and then handle the $N$s accordingly. Still use the example shown in Fig.\ref{fig:LimitationOfOutput2}(b).  If we know, by inferring from the data of PIR sensor, there is a turning point C on the trace A-B, we can use the following equation to estimate the azimuth change $\theta$ during A-B: 
\begin{equation}
\theta \approx |N_1 - N_2| \cdot \theta_c
\label{eqn:modified EST by input}
\end{equation}
where $N_1$ and $N_2$ are the number of the zones crossed before and after turning point C, respectively. 

Eq. \ref{eqn:modified EST by input} can be generalized to  conditions where a trace contains multiple turning points. In these conditions, $M$ turning points will divide the whole trace into $M+1$ sections, and these sections can be divided into two categories. Sections in the same category will have the same direction (clockwise or anti-clockwise). Then we simply let $N_1$ be the number of zones crossed in the first category and  $N_2$ be the number of zones in the second category, and utilize Eq. \ref{eqn:modified EST by input} to obtain the azimuth change during the whole period.

\begin{figure}[h]
\centering
\includegraphics[width=0.6\linewidth]{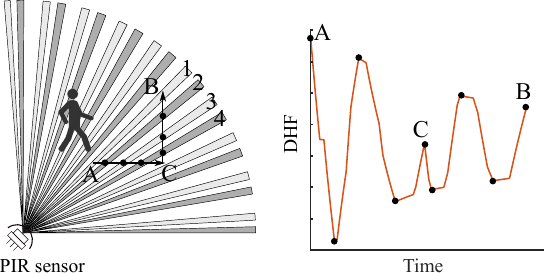}
\caption{Identifying a turning point. (a) A person moves on a trace with a turning point $C$. (b) The corresponding DHF.}
\label{fig:DistinguishTheTurningPoint}
\end{figure}
The last problem is how to identify the changing points. We utilize an example to illustrate the basic idea. Fig.\ref{fig:DistinguishTheTurningPoint} (b) shows the DHF data of a PIR sensor when a person moves along the trace shown in Fig.\ref{fig:DistinguishTheTurningPoint} (a). In the DHF data shown in Fig.\ref{fig:DistinguishTheTurningPoint} (b), there is a local maximum $C$ with a small height. This poiont exactly corresponds to the turning point C in Fig.\ref{fig:DistinguishTheTurningPoint} (a). The reason is as follows. When the person moving on the trace A-C is leaving the negative zone 4, the DHF starts to increase.  However, he makes a turn at point C and then approaching to zone 4 again, and the DHF will then decrease and we can see a local maximum in the DHF data. Likewise, if C is located in the middle of zone 3 and 4, the DHF will has a local minimum. To summarize, a turning point will generate a local extrema in the DHF signal, and we can utilize this character to detect turning points. 

However, as crossing positive/negative zones will also generate local extrema, and we need to distinguish whether a local extrema is caused by turning or by crossing zone.  We utilize a simple threshold method and the threshold to determine heuristically. Through experiment, we determine a peak (or a trough) is caused by turning point if height is smaller than 1/2 of the previous one. 

We should note that, this method can still generate both false negatives (missing identifying turning points)  and false positives (erroneously identifying turning points when there are none). For example, when the turning point happens to be very close to the symmetric axis of a positive/negative zone, we will not observe a local extrema, as it will be merged by the large extrema caused by crossing the symmetric axis itself. This condition will cause false negatives. On the other hand, if a person is walking far away to a PIR sensor without making a turning, it is possible that two peaks have significantly different heights, thus generating false positives.

\section{Localization by azimuth change}
\label{Localization by azimuth change}
\indent In this section, we briefly introduce how to localize a moving person solely by the information of azimuth change. 


We adopt a widely used localization algorithm called as particle filter \cite{cite:An Overview of particle filter} in our application. To use particle filter for localization, we first need to build a \emph{system model}, which describes how the internal states (i.e. location, velocity, etc.) of a moving object change with time, and an \emph{observation model} that describes the relationship between the above internal states and the observed information (i.e. the azimuth change). We utilized a popular system model below:
\begin{equation}
\left[ \begin{array}{l}
{x}_k\\
{y}_k\\
{{\dot x}_k}\\
{{\dot y}_k}
\end{array} \right] = \left[ {\begin{array}{*{20}{c}}
1&0&T&0\\
0&1&0&T\\
0&0&1&0\\
0&0&0&1
\end{array}} \right]\left[ \begin{array}{l}
{x_{k - 1}}\\
{y_{k - 1}}\\
{{\dot x}_{k - 1}}\\
{{\dot y}_{k - 1}}
\end{array} \right] + {\bm{u}_k}
\label{eq:system model}
\end{equation}
where $x_k$ and $y_k$ are the x and y coordinates of a moving person at the $k$th observation time; ${\dot x}_k$ and ${\dot y}_k$ are the corresponding velocities; $T$ is the estimation period; $\bm{u}_k$ is the noise that characterizes the uncertainty in the system model. 
\begin{figure}[h]
\centering
\includegraphics[width=0.45\linewidth]{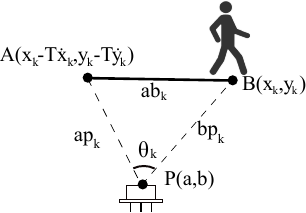}
\caption{Relationship between azimuth change and moving states of a person.}
\label{fig:Relationship between Loaction And DirChange}
\end{figure}

To build the observation model, we find the relationship between the states and the obtained azimuth change. As shown in Fig. \ref{fig:Relationship between Loaction And DirChange}, assume the PIR sensor is located at $(a,b)$, the azimuth change at time $k$ (i.e. $\theta_k$) can be determined by the current state vector $[x_k,y_k,{\dot x}_k,{\dot y}_k]$ through the law of cosines:
\begin{equation}
{\cos\theta_k} =  \frac{{a{p_k}^2 + b{p_k}^2 - a{b_k}^2}}{{2a{p_k} \cdot b{p_k}}}+ {\bm{n}_k}
\label{eq. observation model}
\end{equation}
where $\bm{n}_k$ is the observation noise; $a{p_k}$, $a{b_k}$ and $ab_k$ can be  expressed by the state vector as below:
\[\left\{ \begin{array}{l}
a{p_k} = \sqrt {{{({x_k} - T{{\dot x}_k} - {a})}^2} + {{({y_k} - T{{\dot y}_k} - {b})}^2}} \\
b{p_k} = \sqrt {{{({x_k} - {a})}^2} + {{({y_k} - {b})}^2}} \\
a{b_k} = T \cdot \sqrt {{{\dot x}_k}^2 + {{\dot y}_k}^2} 
\end{array} \right.\]

Eq. \ref{eq. observation model} can be applied to all the deployed PIR sensors and therefore we have a set of equations. This set of equations forms the observation model. Having known the system model and the observation model, we utilize the particle filter framework  \cite{cite:An Overview of particle filter} to determine the person's real-time location.

\section{Experiments}
\label{sec:Experiments}

\subsection {The PIR sensor and its sensing zones}
\label{sec:testlayout}

We adopt PIR sensor Tranesen-PCD-2F21 \cite{cite:handbook of PIR sensors}, and the Fresnel lens array YUYING-8719 \cite{cite:Handbook of Fresnel lens array}. They are off-the-shelf devices and widely used in many PIR applications. 


To estimate the azimuth change, we need to first find out $\theta_c$. At the end of Section 
\ref{sec:Limitations of estimating azimuth change directly by output}, we introduced a Monte Carlo-like method to estimate $\theta_c$.  In this section, we introduce the experimental result to verify the proposed method. 

\begin{figure}[h]%
\centering
\includegraphics[width=0.6\linewidth]{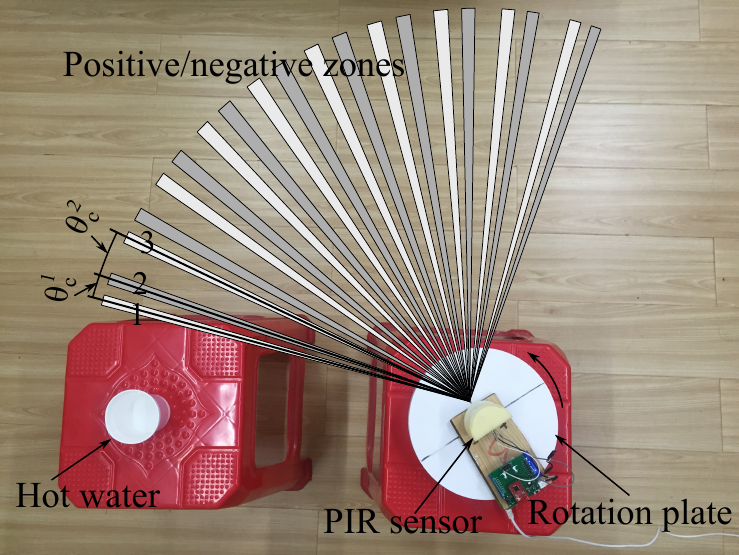}%
\caption{The experimental setup to identify the layout of positive/negative zones of the PIR sensor.}
\label{fig:Verfication}%
\end{figure}

The experimental setup is illustrated in Fig. \ref{fig:Verfication}.  In particular, we placed a bottle of hot water 0.5m away from the PIR sensor as the heat source. The PIR sensor, along with the Fresnel lens array are put on a rotation plate which can rotate at a pre-defined constant speed. During experiment, the rotation plate kept rotating at a constant speed $15^\circ/s$.

\begin{figure}[h]%
\centering
\includegraphics[width=0.65\linewidth]{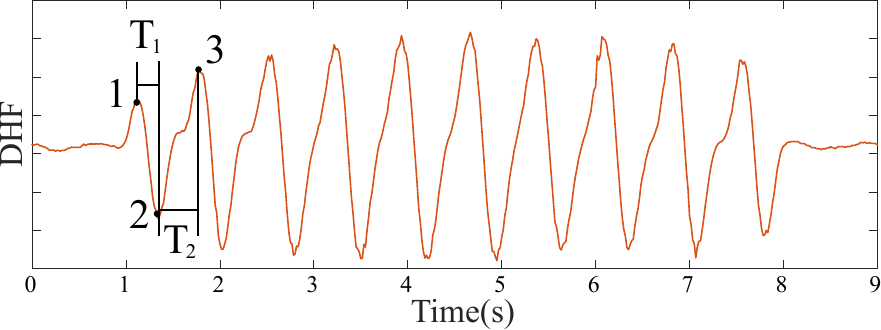}%
\caption{The obtained DHF data and how the peak/troughs correspond to the time instants shown in Fig. \ref{fig:Verfication}. }
\label{fig:DHF_verification}%
\end{figure}

Fig. \ref{fig:DHF_verification} shows a section of the DHF data obtained from the above experiment. We can see that the DHF has 10 peaks and 10 troughs, that corresponds well with the description of the manual that the PIR sensor has 20 zones (10 positive and 10 negative). A peak/trough in the DHF data indicates that the heat source is at the symmetric axis of a positive/negative zone. For example, points 1,2, and 3 in Fig. \ref{fig:DHF_verification} correspond to time instants when the bottle of hot water passes the symmetric axes 1,2, and 3 shown in Fig. \ref{fig:Verfication}, respectively. 

Correspondingly, from the DHF data, we first identify the peaks and troughs, and then calculate time gaps between neighboring peaks and troughs. For example, $T_1$  shown in Fig. \ref{fig:DHF_verification} is the gap between the first peak and first trough, and corresponds to the $\theta_c^1$ shown in Fig. 	\ref{fig:Verfication} based on $
\theta_c^1 = T_1\cdot \omega$
where $\omega$ is the angular speed of the rotation plate. 

\begin{figure}[h]%
\centering
\includegraphics[width=0.7\linewidth]{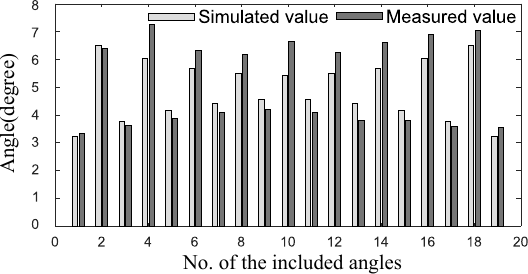}%
\caption{The obtained $\theta_c^i$ ($i=1\sim 19$) and the simulated $\theta_c^i$. }
\label{fig:VerificaitonResult}%
\end{figure}

The equation above holds for all the neighboring zones. Fig. \ref{fig:VerificaitonResult} shows the obtained $\theta_c^i$ ($i=1\sim 19$) and the simulated $\theta_c^i$ introduced in Section 
\ref{sec:Limitations of estimating azimuth change directly by output}. The absolute error of estimated $\theta_c^i$ has mean $0.53^\circ$ and standard deviation $0.35^\circ$.

Furthermore, we take the average of all the obtained $\theta_c^i$, and the final $\theta_c =5.1^\circ$, which is very close to $4.9^\circ$ obtained using the simulation method. The above conclusion demonstrates that using the proposed simulation method, we can accurately estimate the $\theta_c$ of the PIR sensor.

\subsection{Experiments to verify the accuracy of estimated azimuth change}

\begin{figure}[h]
\centering
\includegraphics[width=0.99\linewidth]{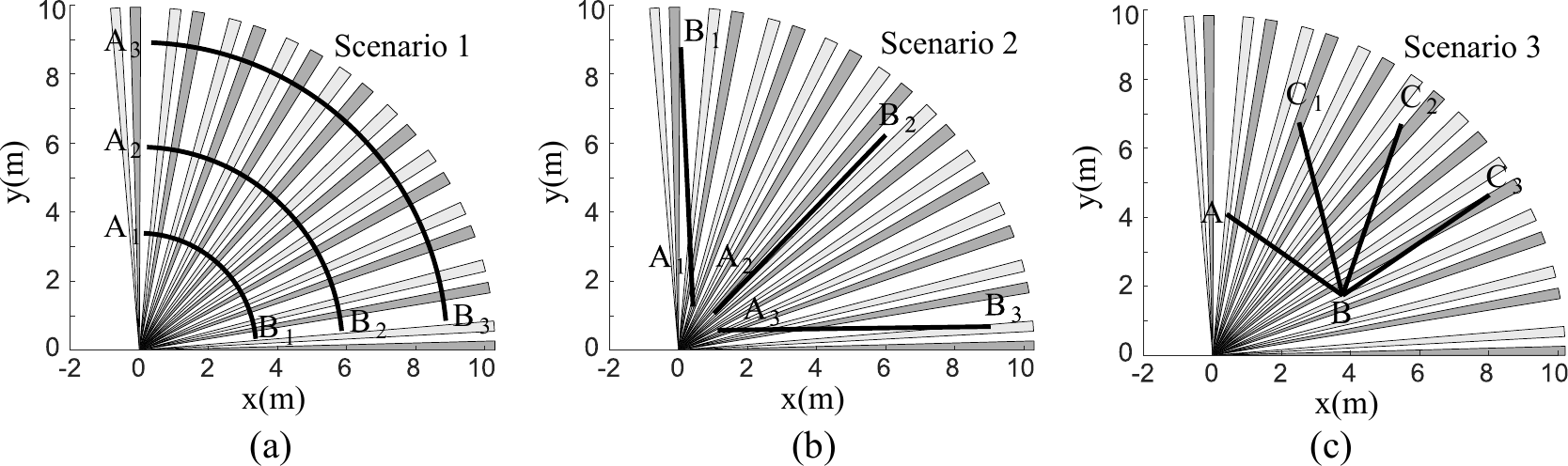}
\caption{Three  scenarios to test the accuracy of the azimuth change estimation.}%
 \label{fig:threescenarios}
\end{figure}

In this section, we introduced the experimental results about the accuracy of the estimated azimuth change. We designed three types of scenarios shown in Fig.	\ref{fig:threescenarios}. In the first scenario (Fig.	\ref{fig:threescenarios}(a)), the person is moving on three arcs, $A_1-B_1, A_2-B_2$ and $A_3-B_3$, all centered at the PIR sensor but with different radius.  We wish to test the performance of the proposed azimuth change estimation method when a person is moving approximately perpendicular to the zones of a PIR sensor. In the second scenario shown in Fig.	\ref{fig:threescenarios}(b), the person is moving on traces approximately parallel to the zones. In the third scenario shown Fig.	\ref{fig:threescenarios}(c), the moving person walks along $A-B$, and then makes turns to $C_1, C_2$ and $C_3$ respectively. We wish to test the performance of the method when the traces contain abrupt turning points. 

\begin{figure}[h]
\centering
\includegraphics[width=0.65\linewidth]{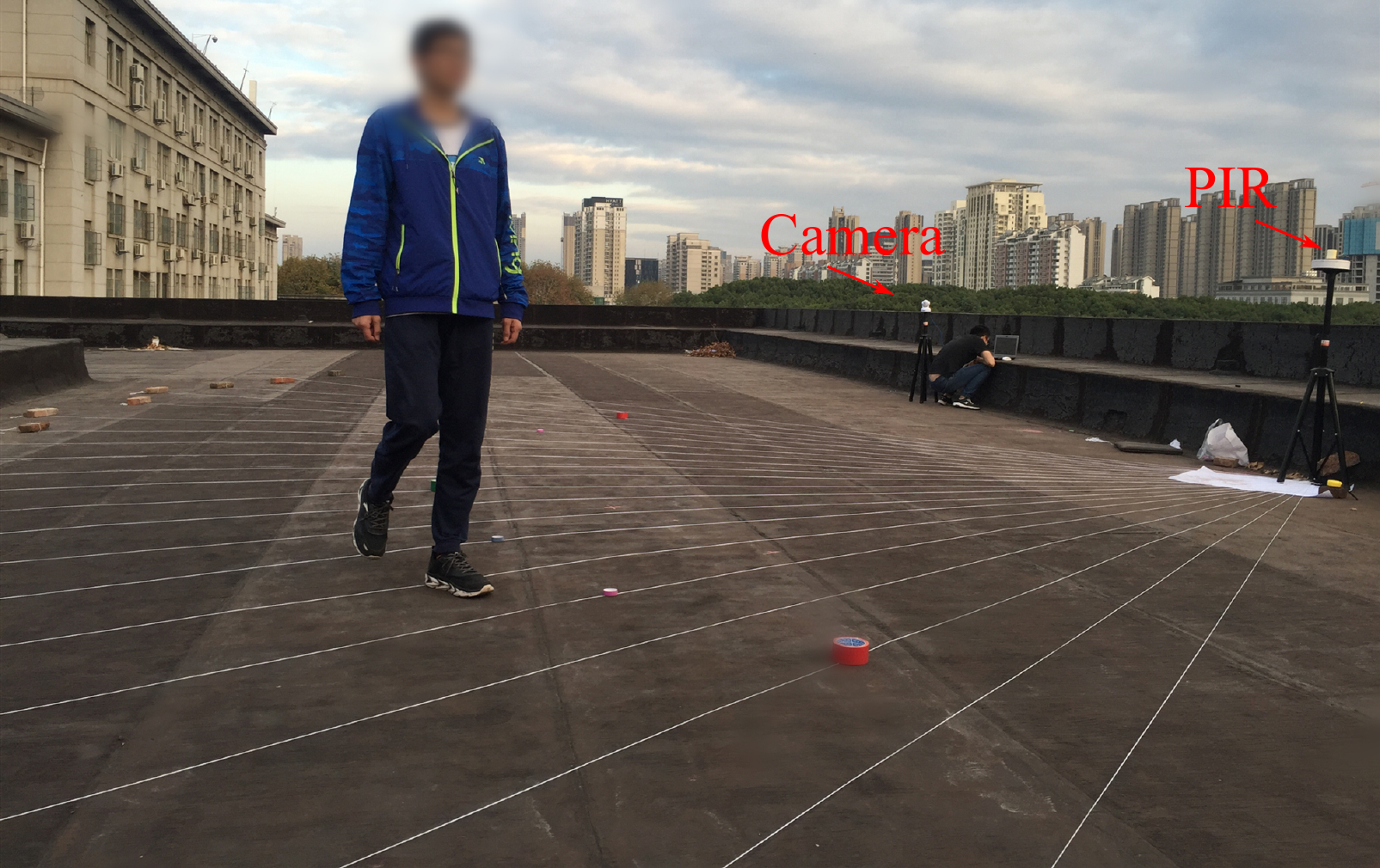}
\caption{Experimental setup.}
\label{fig:setup}
\end{figure}

The experimental setup is shown in Fig. \ref{fig:setup}. In particular, having obtained the layout of the zones of the PIR sensor, we `draw' the layout on the floor using white strings shown in Fig. \ref{fig:setup}. These strings correspond to the edges of the estimated zones. The PIR sensor is aligned with the string-based layout via the rotation plate shown in Fig. \ref{fig:Verfication}. The purpose of establishing such a string-based layout is to guide the person to move along the predefined curves.

We first identify the DHF data from the obtained output of the PIR sensor. Then the corresponding azimuth change in every 0.5s period is estimated. The ground-truth locations are obtained through a camera-based localization system that adopts the pedestrian detection algorithm introduced in \cite{cite:pedestrian detection}.

\begin{figure}[h]%
\centering
\includegraphics[width=0.95\linewidth]{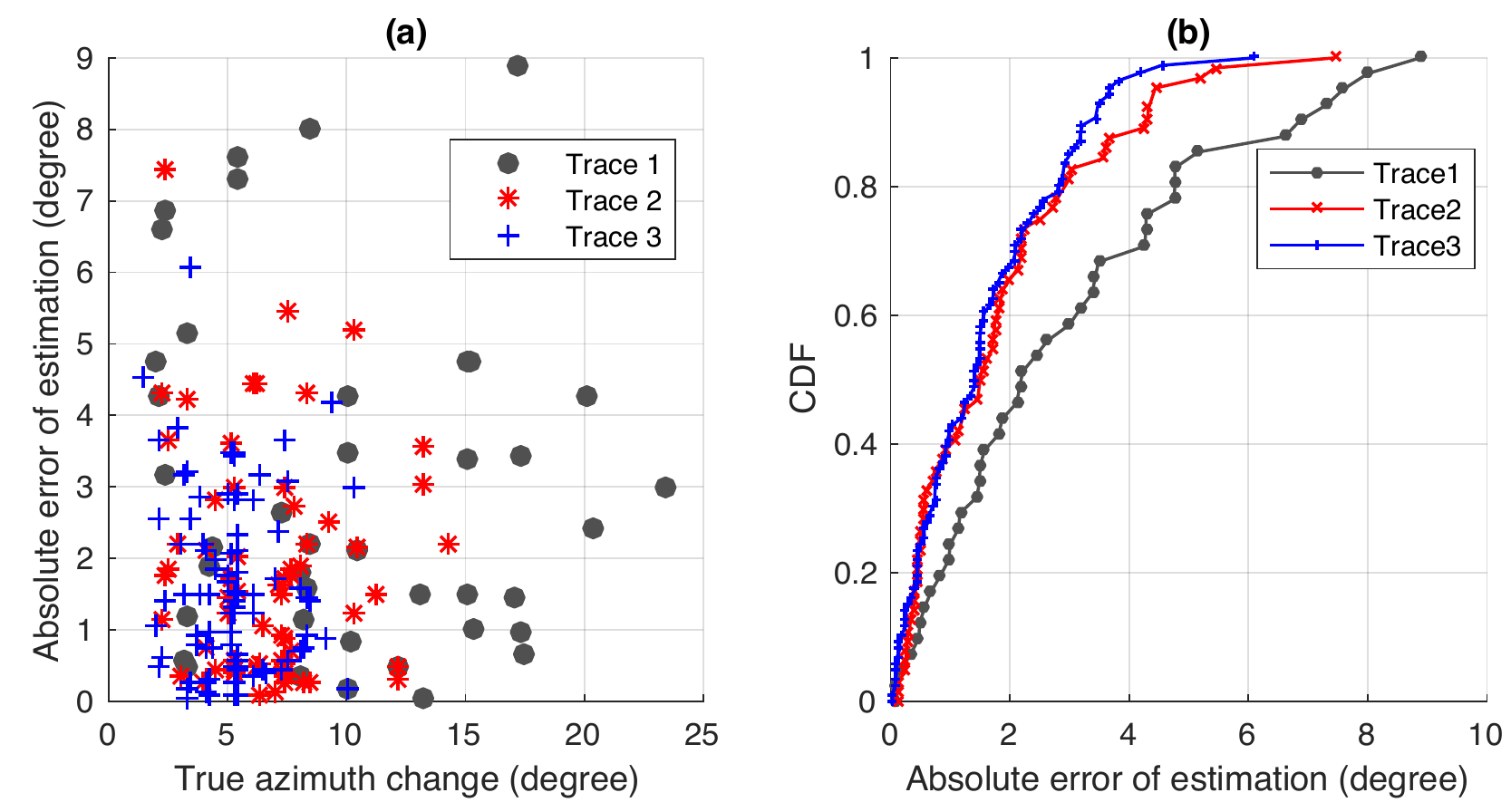}%
\caption{Results in the first scenario. (a) The error of the estimate azimuth change and (b) the corresponding CDF.}
\label{fig:s1}%
\end{figure}

For the first scenario shown in Fig.\ref{fig:threescenarios}(a), the mean and standard deviation (std) of the absolute estimation error for different traces are shown in Table \ref{tab:senario1}.  
\begin{table}[h]
\small
\centering
\begin{tabular}{| c | c | c | c |}
	\hline
	& Trace 1 & Trace 2 & Trace 3 \\
	&($A_1-B_1$) & ($A_2-B_2$) & ($A_3-B_3$)\\ \hline
	Mean (degree) & 3.0 & 1.8 & 1.5 \\ \hline
	Std (degree) & 2.3 & 1.5 & 1.2 \\ \hline
\end{tabular}
\caption{The statistics of the three traces in scenario 1.}
\label{tab:senario1}
\end{table}
The detailed results are shown in Fig.\ref{fig:s1}. In particular, the x-axis of Fig.\ref{fig:s1}(a) is the real azimuth change obtained from the camera-based system, and the y-axis is the absolute error of the estimated ones. We can see that the inner arc $A_1-B_1$ has the largest error while the outer arc $A_3-B_3$ has the smallest one. This can be attributed to fact that the size of zones near the PIR sensor is much smaller than that of zones far away from the sensor. When a person is very close to the PIR sensor, the body may within multiple zones simultaneously at any moment, and therefore the DHF data do not show clear peaks/troughs, leading to a larger estimation error. Fig. \ref{fig:s1}(b) shows the corresponding cumulative distribution function (CDF) of the error of the estimated azimuth change.

\begin{figure}[h]%
\centering
\includegraphics[width=0.95\linewidth]{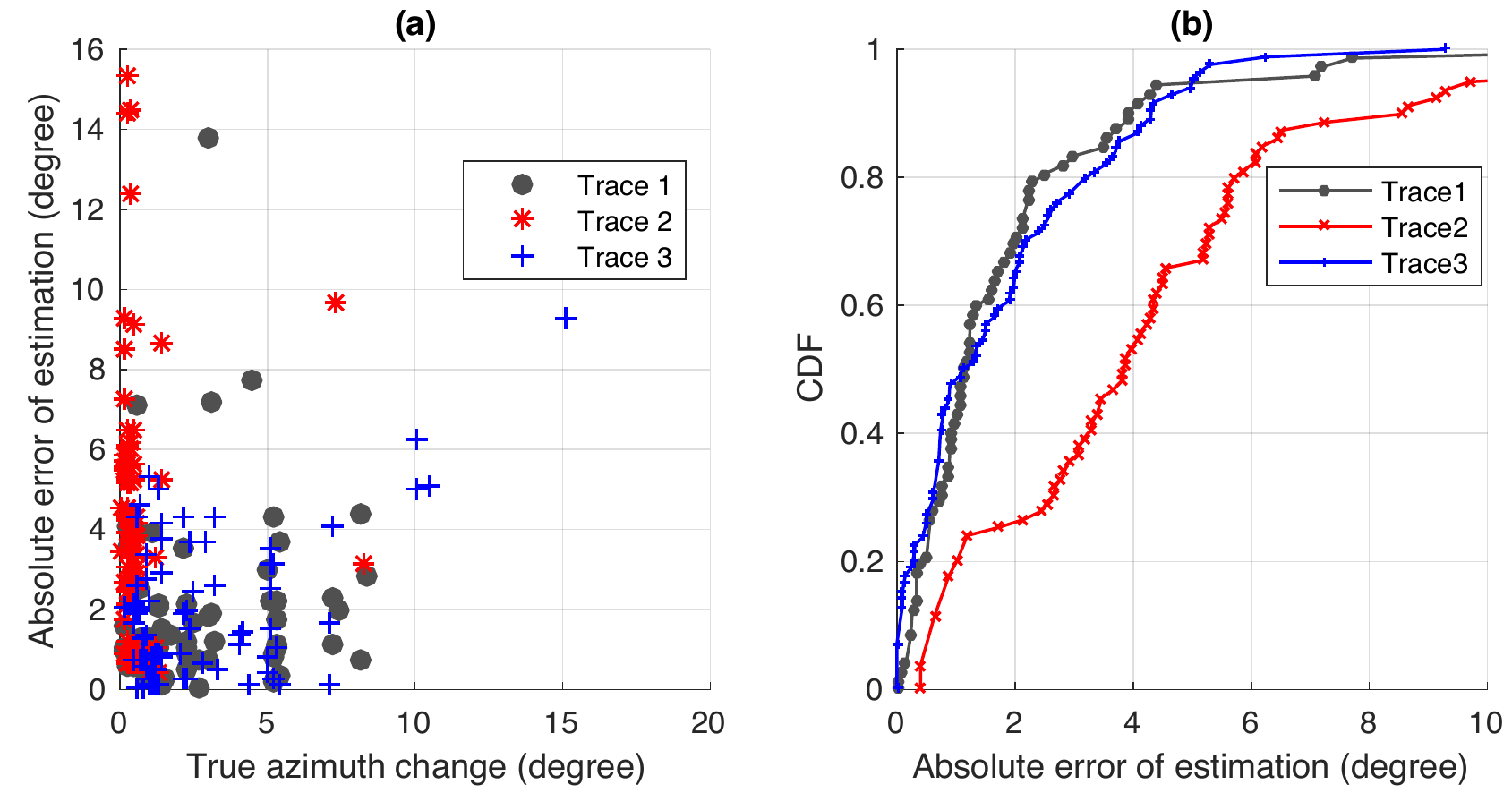}%
\caption{Results in the second scenario. (a) The error of the estimate azimuth change and (b) the corresponding CDF.}
\label{fig:s2statistics}%
\end{figure}

For the second scenario shown in Fig.\ref{fig:threescenarios}(b), Table \ref{tab:senario2} shows the mean and std of the absolute estimation error. The details are illustrated in Fig. \ref{fig:s2statistics}. We can see that trace 2 (i.e. $A_2-B_2$) has much higher error the other two. By examining the traces in Fig.\ref{fig:threescenarios}(b), we can see that trace 2 is almost parallel to a zone of the PIR sensor, and the person barely crossed a complete zone on the whole trace. Correspondingly, the true azimuth change remains to be close to $0^\circ$. As no zones have been really passed, the DHF data oscillate slightly, and we potentially have larger estimation error. The person on  trace 1 (i.e. $A_1-B_1$) and trace 3 (i.e. $A_3-B_3$) did crossed a few zones, and therefore have smaller estimation errors. 

\begin{table}
\centering
\small
\begin{tabular}{| c | c | c | c |}
	\hline
	& Trace 1 & Trace 2 & Trace 3 \\
	&($A_1-B_1$) & ($A_2-B_2$) & ($A_3-B_3$)\\ \hline
	Mean (degree)& 1.8  & 4.2 & 1.7 \\ \hline
	Std (degree) & 2.1 & 3.2 & 1.7 \\ \hline
\end{tabular}
\caption{The statistics of the three traces in the scenario 2.}
\label{tab:senario2}
\end{table}

\begin{figure}[h]%
\centering
\includegraphics[width=0.95\linewidth]{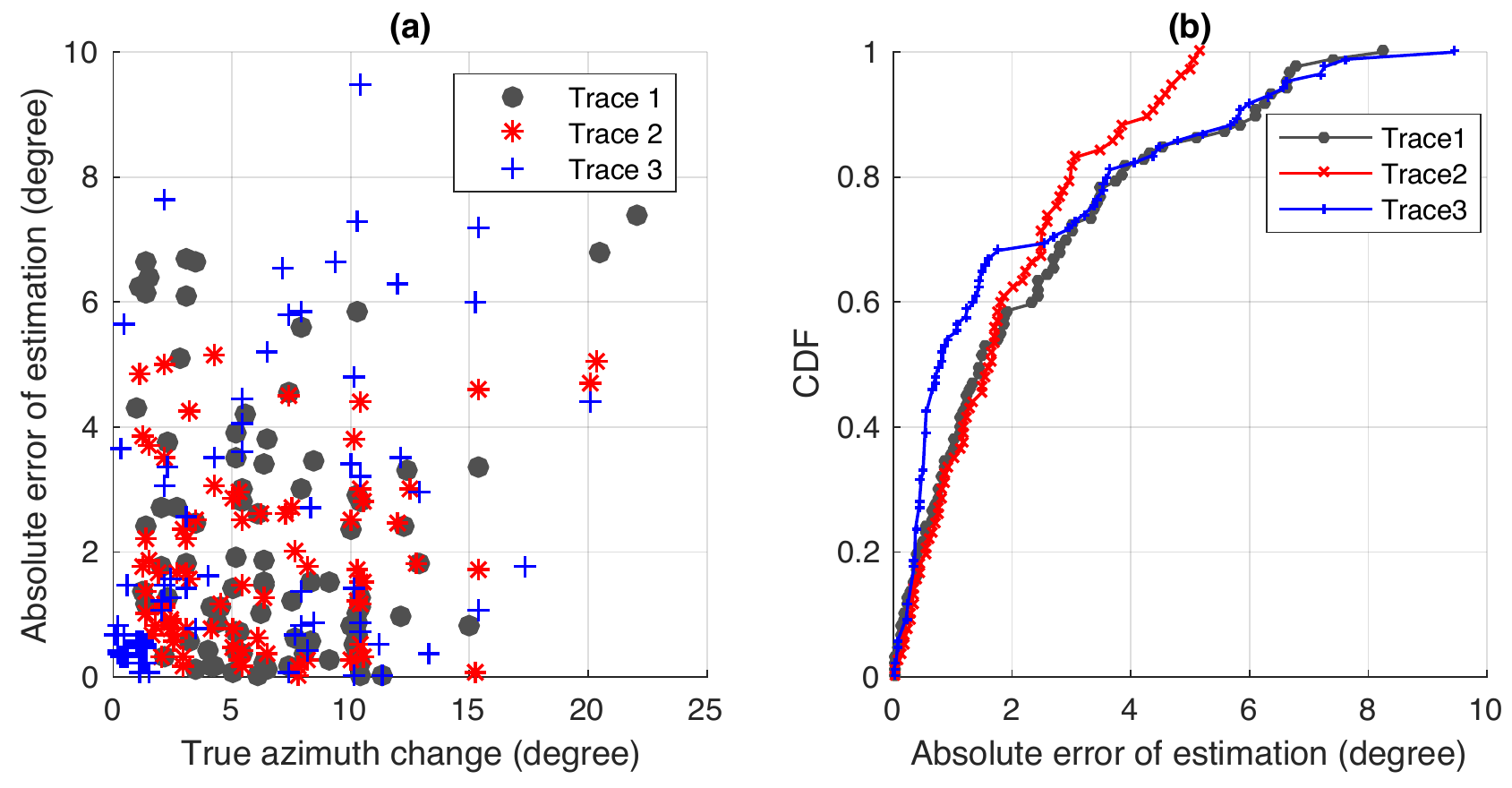}%
\caption{Results in the third scenario. (a) The error of the estimate azimuth change and (b) the corresponding CDF.}
\label{fig:s3}%
\end{figure}

Table \ref{tab:senario3} shows statistics of the absolute estimation error in the third scenario shown in Fig.\ref{fig:threescenarios}(c). The details are illustrated in Fig. \ref{fig:s3}. We can see that in traces containing abrupt turning points, the obtained estimation error has the same level with those without such turning points.

\begin{table}
\centering
\small
\begin{tabular}{| c | c | c | c |}
	\hline
	& Trace 1 & Trace 2 & Trace 3 \\
	&($A-B-C_1$) & ($A-B-C_2$) & ($A-B-C_3$)\\ \hline
	Mean (degree) & 2.2  & 1.9 & 1.9 \\ \hline
	Std (degree) & 2.1 & 1.4 & 2.2 \\ \hline
\end{tabular}
\caption{The statistics of the three traces in scenario 3.}
\label{tab:senario3}
\end{table}

\subsection{Experiments to verify the accuracy of localization}
\label{sec:Tracking performance}

In this section, we demonstrate the performance of a PIR-based localization system solely based on the azimuth change.

\begin{figure}
\includegraphics[width=0.8\linewidth]{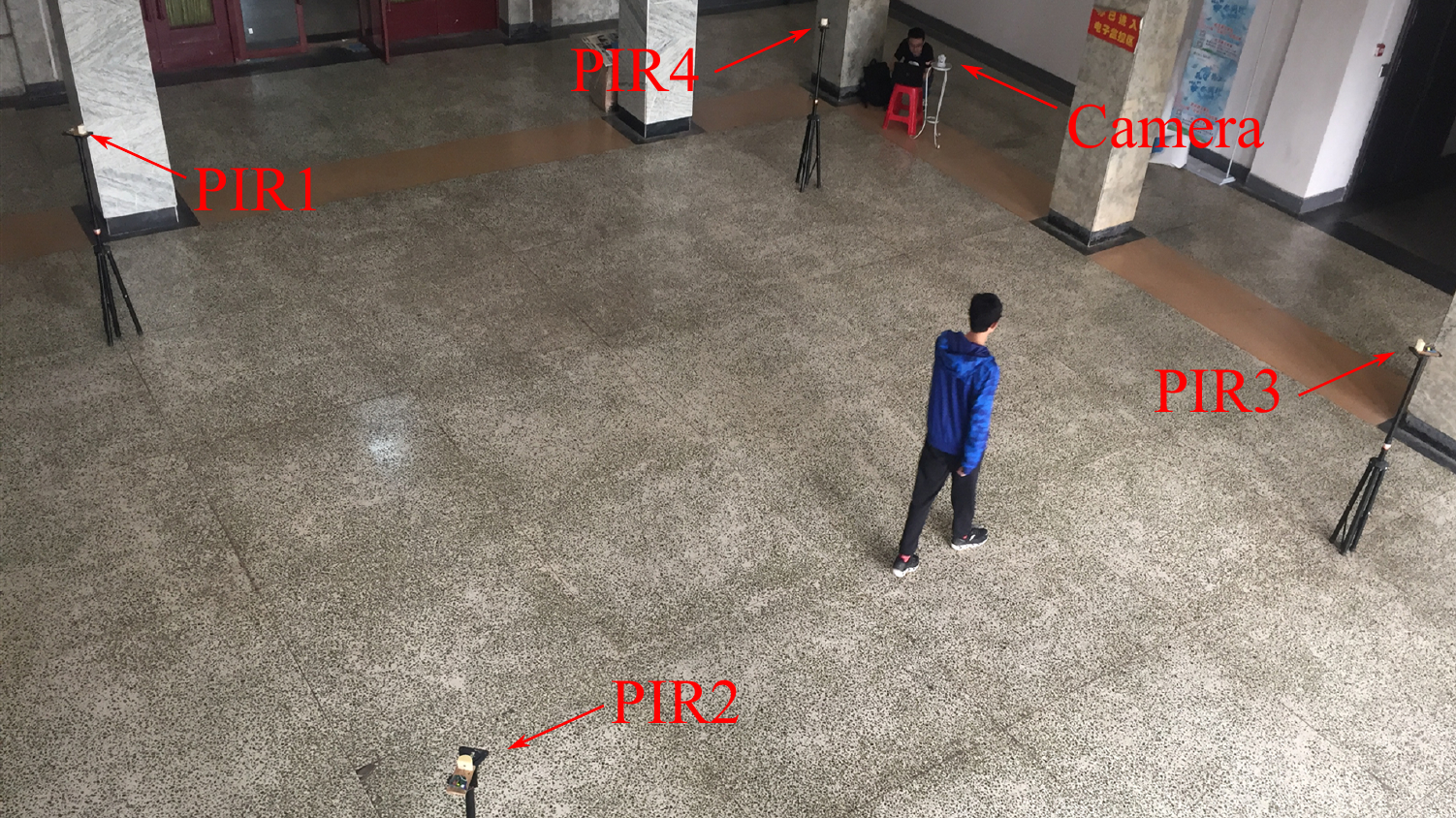}%
\caption{The setup of a PIR-based localization system.}
\label{fig:localization}%
\end{figure}

The setup of the localization system is shown in Fig. \ref{fig:localization}. The system contains 4 PIR sensors located at corners of a $7m-by-7m$ area. In addition, the output of the PIR sensor are transmitted through a wireless communication system (consisting of five CC2530 devices and a CP2102 model) to a host computer, where the DHF data are estimated to obtain the azimuth change. The estimated azimuth change is fed into the particle filter to obtain the real-time location.

\begin{figure}
\centering
\includegraphics[width=0.9\linewidth]{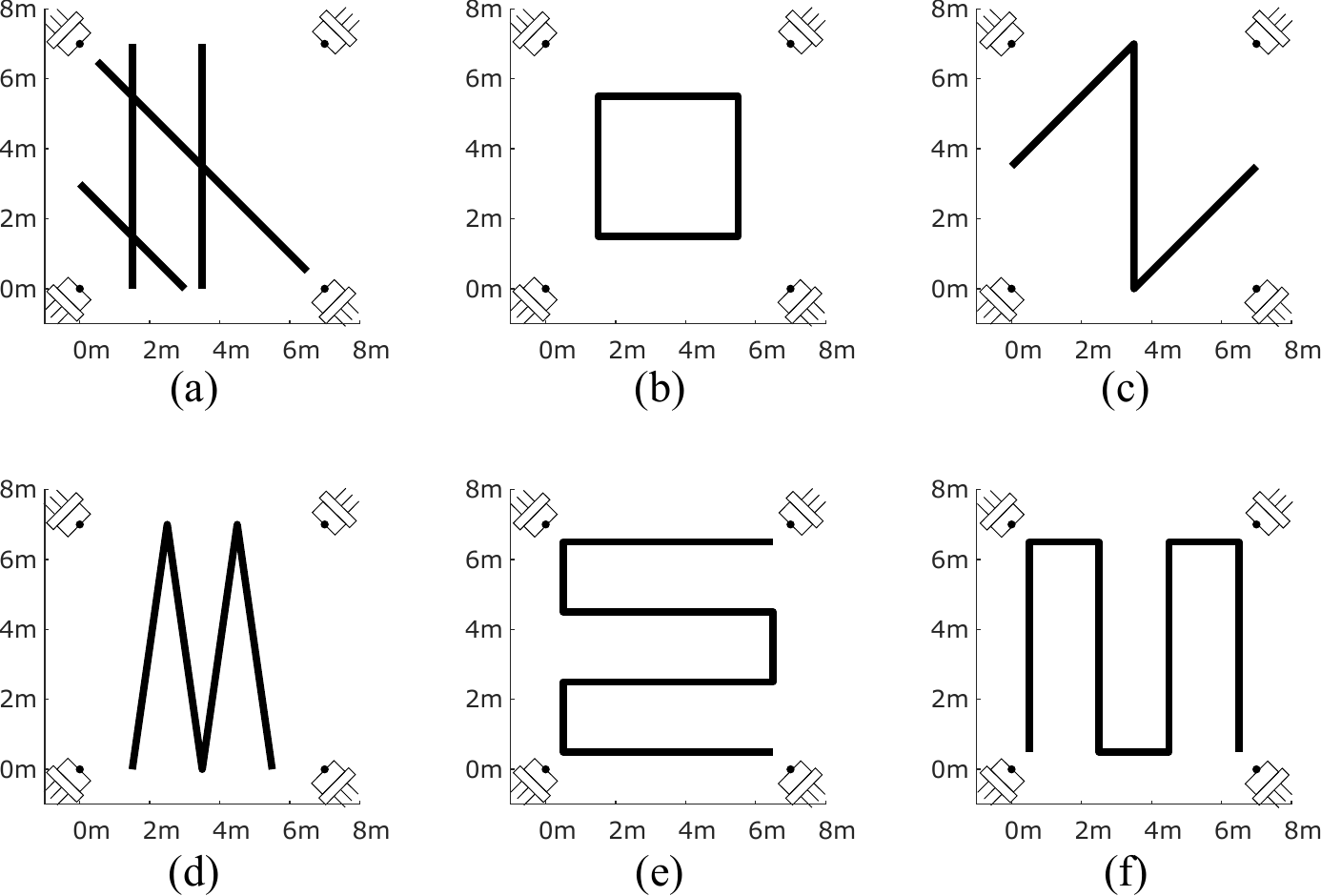}%
\caption{Six testing scenarios. (a) Straight lines, (b) a square, (c) Z shape, (c) M shape, (d) horizontal snake-like curve, and (e) vertical snake-like curve.}
\label{fig:6scenrios}%
\end{figure}

We designed a total of 6 testing scenarios shown in Fig. \ref{fig:6scenrios}. These scenarios contain simple straight lines (Fig.\ref{fig:6scenrios}(a))  or more complex curves.  In each scenario, the person was arranged to take a round trip five times along the pre-defined traces.

\begin{figure}
\centering
\includegraphics[width=0.8\linewidth]{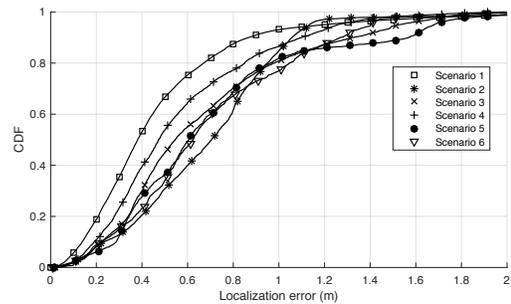}%
\caption{The CDF of the localization error in 6 scenarios.}
\label{fig:CDFoflocalization}%
\end{figure}

Table \ref{tab:6senario} gives the statistics of the localization error in terms of mean and standard deviation. The detailed CDF of the localization error in each of the 6 scenarios is shown in Fig. \ref{fig:CDFoflocalization}.

\begin{table}
\centering
\small
\begin{tabular}{|l|*{6}{c|}}\hline
	\backslashbox{Error}{Scenario}
	&\makebox[2em]{1}&\makebox[2em]{2}&\makebox[2em]{3}
	&\makebox[2em]{4}&\makebox[2em]{5}&\makebox[2em]{6}\\ \hline
	Mean (m) &0.47 & 0.69 & 0.66 & 0.56 & 0.71& 0.69\\ \hline
	Std (m) & 0.41& 0.37 & 0.41 & 0.35 & 0.45& 0.38\\ \hline
\end{tabular}
\caption{The statistics of the localization error in 6 scenarios.}
\label{tab:6senario}
\end{table}

We can see that our PIR-based localization can achieve sub-meter localization accuracy with $80\%$ or higher probability in all scenarios. In particular, the system has the highest performance in scenario 1 (with sub-meter accuracy in $92\%$). The average localization error of all scenarios is about $0.63$m with standard deviation $0.39$m.  Even the worst scenario can achieve localization error smaller than $1.8$m with $99\%$ probability. The results demonstrate the effectiveness of using the azimuth change for localization.


\textbf{The effect of estimation period}

\begin{figure}
\centering
\includegraphics[width=0.7\linewidth]{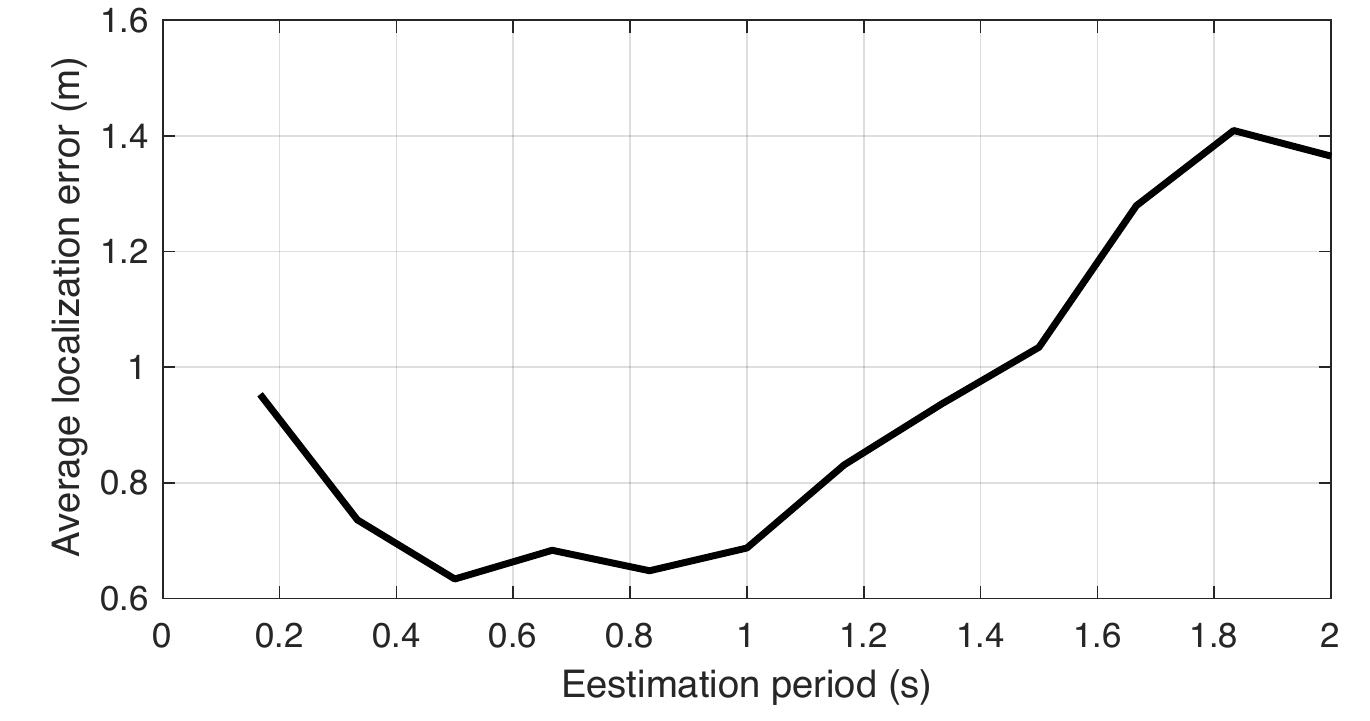}%
\caption{The effect of the estimation period.}
\label{fig:effectofestimationperiod}
\end{figure}

In Section \ref{sec:Influence of repeatedly crossed fan-shaped zones}, we mentioned that it is important to choose a proper estimation period.  Fig. \ref{fig:effectofestimationperiod} shows, using data from 6 scenarios above, how the average localization error changes with different estimation period.  We can clearly observe a `U' curve when the estimation period changes from 0.1s to 2s. A very small estimation period like $0.1$s will lead to a relatively large localization error because in such a short period of time, the moving person is rarely able to cross a complete zone of the PIR sensor. On the other hand, a large estimation period like $2$s may contain many abrupt turning points. These abrupt turning points, although can be mitigated by the method proposed in Section \ref{sec:Influence of repeatedly crossed fan-shaped zones}, their effect cannot be  completely removed since not all the abrupt turning points can be successfully detected.  The localization error reaches its minimum about $0.5s\sim 1s$. This justifies our setting of the estimation period $0.5$s in previous experiments.

\textbf{The effect of number of PIR sensors}

\begin{figure}
\includegraphics[width=0.99\linewidth]{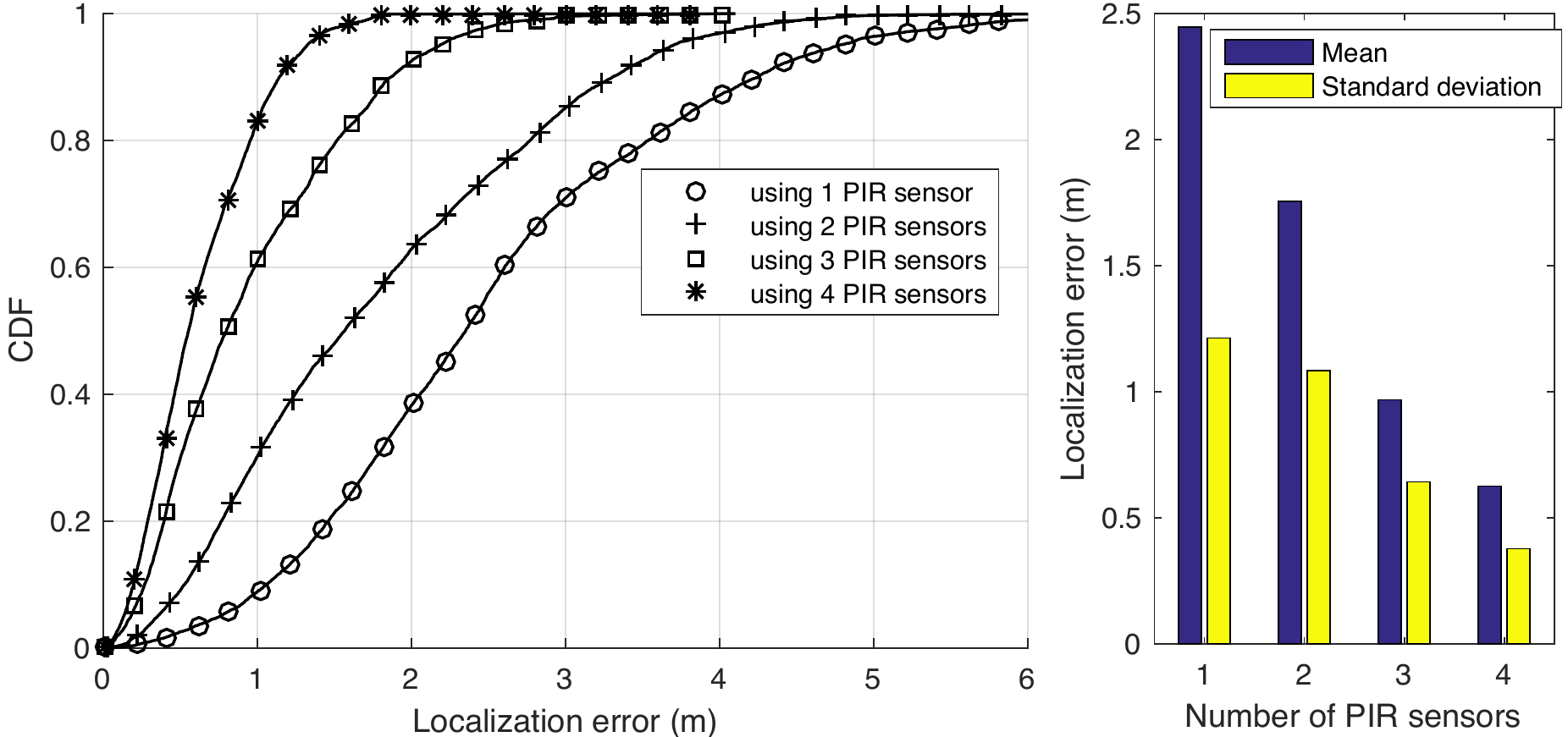}%
\caption{The CDF of localization error for systems with different number of PIR sensors.}
\label{fig:thenumberofPIRsensor}%
\end{figure}

We also test the effect of the number of PIR sensors on the accuracy of the localization system. Fig. \ref{fig:thenumberofPIRsensor}(a) shows the CDF of localization error for systems with different number of PIR sensors. The corresponding mean and standard deviation are shown in Fig. \ref{fig:thenumberofPIRsensor}(b). We can see that the average localization error is $2.5$m for a localization system with a single PIR sensor but drops to $1.7$m, $0.9$m, and further down to $0.63$m if more PIR sensors are added.

\textbf{Comparing to the PIR-based localization system \cite{cite:PIR Sensors Characterization and Novel Localization Technique}}

We  compare the localization accuracy of our system to the one proposed in \cite{cite:PIR Sensors Characterization and Novel Localization Technique}, which utilizes a data-driven approach to establish a model describing the relationship between the amplitude of PIR's outputs and its distance to a moving person.

\begin{figure}[h]%
\centering
\subfloat[][]{\centering
	\includegraphics[width=0.31\linewidth]{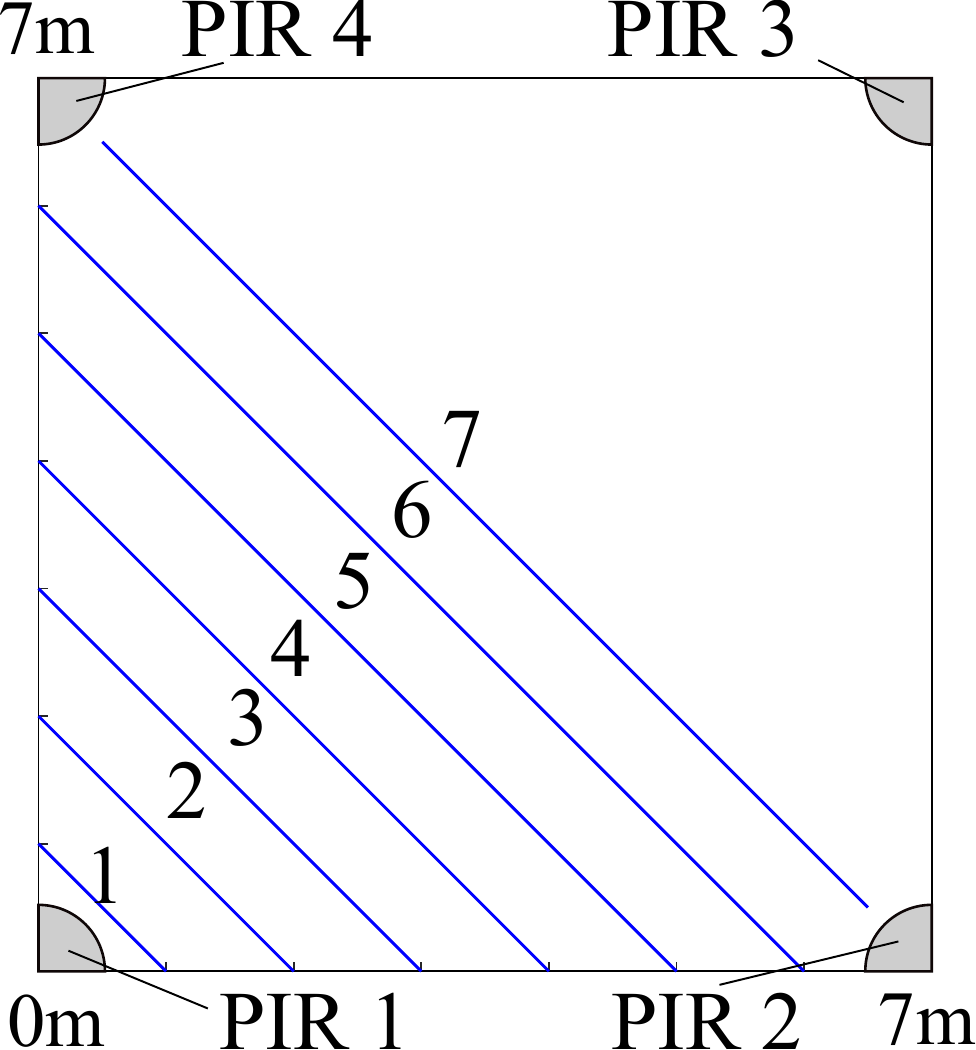}}
\subfloat[][]{\centering
	\includegraphics[width=0.31\linewidth]{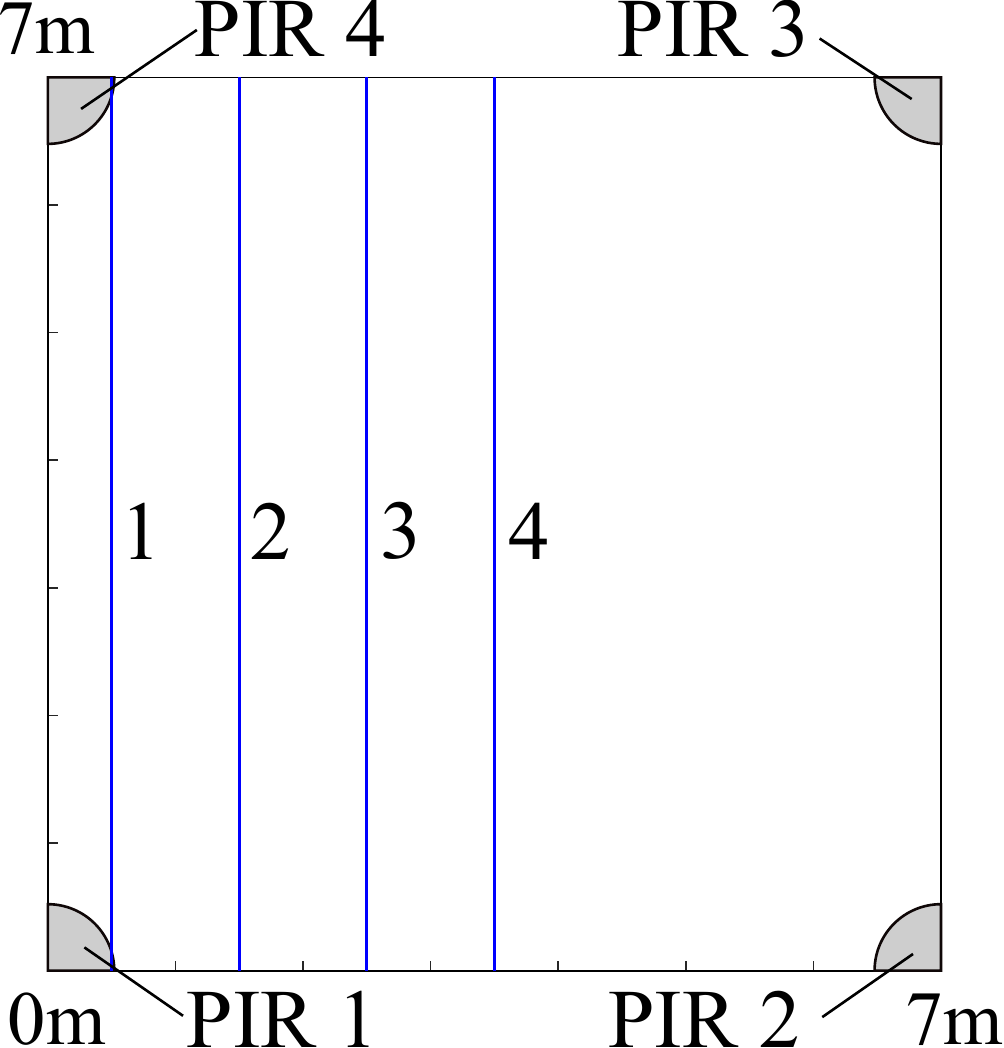}}
  \subfloat[][]{\centering
  \includegraphics[width=0.35\linewidth]{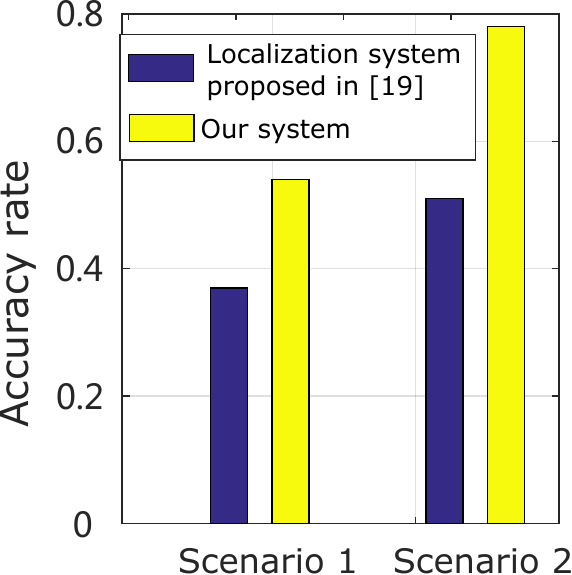}}
\caption[]{Experimental scenarios utilized in \cite{cite:PIR Sensors Characterization and Novel Localization Technique} and comparison result.
	(a) 7 slashed parallel traces, (b) 4 vertical parallel traces, (c) comparison results.}%
\label{fig:scenarios of experiment for localization performance verification}%
\end{figure}

For a fair comparison, we use the same traces described in \cite{cite:PIR Sensors Characterization and Novel Localization Technique} shown in Fig. \ref{fig:scenarios of experiment for localization performance verification} (a) and (b).  Furthermore, we utilize the same performance measure in \cite{cite:PIR Sensors Characterization and Novel Localization Technique} called as `accuracy rate', which is the probability of correctly localizing a person within a 1m$\times$ 1m grid.

The comparison results in the two scenarios are shown in Fig.\ref{fig:scenarios of experiment for localization performance verification}(c). It can be seen that in both scenarios, our localization system can achieve better accuracy rate than the system described in \cite{cite:PIR Sensors Characterization and Novel Localization Technique}.  The accuracy rate of our system is 0.54 and 0.78, about $50\%$ improvement on the system proposed in  \cite{cite:PIR Sensors Characterization and Novel Localization Technique}. In addition, considering that the system proposed in \cite{cite:PIR Sensors Characterization and Novel Localization Technique} has 8 PIR sensors and requires abundant training data, while our system only utilizes 4 PIR sensors without collecting training data, the advantage of using azimuth change becomes apparent.

\section{Related works}
\label{related works}

Using PIR sensors to localize one's location has received much attention recently and many PIR-based localization systems have been designed \cite{cite:Human Tracking With Wireless Distributed Pyroelectric Sensors} -\cite{cite:INDOOR USER ZONING AND TRACKING}. Most of the existing PIR-based localization systems take a PIR sensor as a `binary' indicator. Based on binary outputs of multiple PIR sensors, a moving person's can be tracked. 

According to the number of persons that can be simultaneously tracked, the works \cite{cite:Human Tracking With Wireless Distributed Pyroelectric Sensors} and \cite{cite: Moving Targets Detection and Localization in Passive Infrared Sensor Networks} introduced how to determine a single person's real-time location. On the other hand, the works \cite{cite:Distributed Multiple Human Tracking with Wireless Binary PIR Sensor Networks}-\cite{cite:A novel multi-human location method} introduced how to track multiple persons. The relationship between the deployment and the localization accuracy is discussed in \cite{cite:Surveillance Tracking System Using}-\cite{cite:Preprocessing Design in Pyroelectric}. In addition, the works \cite{cite:MOLTS} and \cite{cite:last binary work} discussed how to integrate the PIR sensor's binary information with other kinds of sensors for localization. 


One common limitation of the `binary-based' systems above is that the localization accuracy is mainly depended on the size of the overlapped sensing zones. To achieve high localization accuracy, a large number of PIR sensors will be required. 

Using finer grained information rather than the binary information is a new trend in of designing PIR-based localization systems. To the best of our knowledge, there are only two related works. In \cite{cite:INDOOR USER ZONING AND TRACKING} and \cite{cite:PIR Sensors Characterization and Novel Localization Technique},  the raw output data of PIR sensors are utilized to localize a person. However, there are still limitations of the two works above. First, they are both data-driven and require abundant of training data to be collected in different environments and different types of PIR sensors. 



\section{Discussions and conclusion}

The idea of this paper is simple: from the raw output of PIR sensors, we can extract a type of information called as azimuth change, and this information can be utilized to track a moving person.  In particular, based on the analysis of the physical properties and internal structure of the PIR sensor and Fresnel lens array, we propose an approach to  estimate the azimuth change using the output data of a PIR sensor.  In addition, based on the estimated azimuth change, we built a practical PIR-based localization system which can achieve better performance than the state-of-art work.

There still remain some limitations of using the proposed approach. First, when a person is very close to a PIR sensor, the accuracy of estimated azimuth change will be relatively low. This has been revealed in Fig. \ref{fig:s1}. Although this problem can be partially alleviated through the deployment of PIR nodes, we believe a depth study on this problem can help to improve the localization accuracy of the whole system. 

Another limitation is the system's robustness to environmental noise, especially in an open environment with many static or moving heat sources. This is an intrinsic challenge for a PIR-based localization system. For our system, it is better to design a dynamic threshold to distinguish whether a peak/trough is caused by the moving target or by other heat sources. 

Last but not least, the current method proposed in this paper can only localize a single person. In the future, we will go on exploring how to expand this method for multiple
person localization.



\end{document}